\documentclass[useAMS,usenatbib]{mn2e}

\usepackage{amsmath}
\usepackage{amssymb}
\usepackage{multicol}
\usepackage{rotating}
\usepackage{lscape}

\usepackage{graphicx}
\usepackage{color}
\usepackage{placeins}

\newcommand{\beq}{\begin{equation}}
\newcommand{\eeq}{\end{equation}}
\newcommand{\beqa}{\begin{eqnarray}}
\newcommand{\eeqa}{\end{eqnarray}}

\definecolor{gray}{gray}{0.55}
\def\d{\delta}

\def\del{\nabla}

\font\BF=cmmib10

\def\x{{\hbox{\BF x}}}

\def\q{{\hbox{\BF q}}}

\def\Mpc{\, h^{-1} \, {\rm Mpc}}

\begin{document}

%
\title[A Large Sample of Mock Galaxy Catalogues for BOSS]{
The clustering of galaxies in the SDSS-III Baryon Oscillation Spectroscopic Survey: a large sample of mock galaxy catalogues}

\author[Manera et al. ]{\parbox{\textwidth}{Marc Manera$^{1}$\thanks{email:marc.manera@port.ac.uk},
Roman Scoccimarro$^{2}$, 
Will J. Percival$^{1}$, 
Lado Samushia$^{1}$, 
Cameron K. McBride$^{3}$, 
Ashley J. Ross$^{1}$, 
Ravi K. Sheth$^{4}$$^{,5}$, 
Martin White$^{6}$$^{,7}$,
Beth A. Reid$^{7,8}$,
Ariel G. S\'{a}nchez$^{9}$,
Roland de Putter$^{10}$$^{,11}$, 
Xiaoying Xu$^{12}$,
Andreas A. Berlind$^{13}$, 
Jonathan Brinkmann$^{14}$,
Bob Nichol$^{1}$,
Francesco Montesano$^{9}$,
Nikhil Padmanabhan$^{15}$,
Ramin A. Skibba$^{12}$,
Rita Tojeiro$^{1}$,
and 
Benjamin A. Weaver$^{2}$
} 
\vspace*{2pt}\\
$^{1}$Institute of Cosmology and Gravitation, Portsmouth University, Dennis Sciama Building, Po1 3FX, Portsmouth, UK\\
$^{2}$Center for Cosmology and Particle Physics, New York University, 4 Washington Place, NY 1003, New York, USA\\
$^{3}$Harvard-Smithsonian Center for Astrophysics, 60 Garden St., Cambridge, MA 02138, USA\\
$^{4}$Abdus Salam International Center for Theoretical Physics, Strada Costiera 11, 34151, Trieste, Italy\\
$^{5}$University of Pennsylvania, 209 S. 33rd Street, Philadelphia, PA 19104, USA\\
$^{6}$Departments of Physics and Astronomy, University of California, Berkeley, CA 94720, USA\\
$^{7}$Lawrence Berkeley National Laboratory, 1 Cyclotron Road, Berkeley, CA 94720, USA\\
$^{8}$Hubble Fellow\\
$^{9}$Max-Planck-Insitut f\"{u}r Extraterrestrische Physik, Giessenbachstraße, 85748 Garching, Germany\\
$^{10}$Instituto de Fisica Corpuscular, Universidad de Valencia-CSIC, Spain\\ 
$^{11}$Institut de Ciències del Cosmos, University of Barcelona (IEEC-UB), Marti i Franques 1, Barcelona 08028, Spain\\
$^{12}$Steward Observatory, University of Arizona, 933 North Cherry Ave., Tucson, AZ 85721, USA\\
$^{13}$Department of Physics and Astronomy, Vanderbilt University, Nashville, TN 37235, USA\\
$^{14}$Apache Point Observatory, 2001 Apache Point Road, Sunspot, NM 88349, USA\\
$^{15}$Department of Physics, Yale University, 260 Whitney Ave, New Haven, CT 06520, USA\\ 
}

\maketitle
%
\begin{abstract}
%
We present a fast method of producing mock galaxy catalogues that can
 be used to compute covariance matrices of large-scale clustering measurements and test
 the methods of analysis. Our method populates a 2nd-order Lagrangian Perturbation Theory
 (2LPT) matter field, where we calibrate masses of dark matter halos by detailed
 comparisons with N-body simulations. We demonstrate the clustering of halos is recovered
 at $\sim$10 per cent accuracy. We populate halos with mock galaxies using a Halo
 Occupation Distribution (HOD) prescription, which has been calibrated to reproduce the clustering measurements on scales between  30 and 80 $\Mpc$. We compare the sample covariance matrix from our
mocks with analytic estimates, and discuss differences.  We have used this method to
make catalogues corresponding to Data Release 9 of the Baryon Oscillation Spectroscopic Survey (BOSS),
 producing 600 mock catalogues of the ``CMASS'' galaxy sample. These mocks enabled detailed
tests of methods and errors that formed an integral part of companion analyses of these galaxy data.
\end{abstract}


\section{Introduction}

Galaxy surveys such as the the Baryon Oscillation Spectroscopic 
Survey (BOSS, \citealt{Boss2,Eis11}), WiggleZ \citep{Drinkwater10}, 
HETDEX \cite{HETDEX}, and the
Dark Energy Survey\long\def\symbolfootnote[#1]#2{\begingroup%
  \def\thefootnote{\fnsymbol{footnote}}\footnote[#1]{#2}\endgroup}
\symbolfootnote[2]{http://www.darkenergysurvey.org}, 
designed to cover
large areas of the sky, are currently leading the effort to constrain
cosmological parameters using the observed clustering of galaxies and quasars.  In
future, the baton will be passed to projects such as eBOSS, BigBOSS,
Euclid \citep{Euclid} and LSST \citep{LSST}.  These projects will 
cover large volumes of the Universe, and observe millions of
galaxies in order to make precise measurements.  BOSS
aims to determine the cosmic expansion rate $H(z)$ with a precision of
1 per cent at redshifts $z\simeq0.3$ and $z\simeq0.6$, and with 1.5 per cent at
$z\approx2.5$, by means of accurately measuring the scale of the
baryon acoustic peak \citep{Eis11}.
The first steps towards this goal are presented in a companion paper \citep{boss12}, which 
provides the highest precision measurement of the baryon acoustic scale to date. 

Such large-scale clustering
measurements require an estimate of their covariance matrix in order to produce reliable cosmological constraints. One could
get this matrix by running a large number of N-body simulations and
generating galaxy mocks.  However this is computationally very
expensive and, as surveys probe increasingly larger scales, impractical.  If only a small number of realizations is used
($\approx$50 simulations), then the estimated covariance
matrix can be very noisy. There have been several suggestions in the literature on how to deal with this problem.

When analysing SDSS-II DR7 Luminous Red Galaxies, \cite{Xu12} used a smooth approximation to the mock covariance matrix. This technique involves fitting an analytic form to a covariance matrix that is 
computed from a relatively small number of mock catalogues, using a
maximum likelihood approach with a number of underlying assumptions. 
This smoothing technique is critical in the regime of a small number of mocks, but would be obsolete if a sufficiently large number of mocks were available, requiring fewer underlying assumptions in the estimation of the covariance matrix. Such techniques may also be able to help translate matrices between cosmological models.

Alternatively, the lognormal model has been used to generate large numbers of
mock catalogues, from which covariance matrices are calculated
\citep{Cole05,Percival10,Blake11}.  Because of its simplicity this approach
is fast.  However it does not properly account for non-Gaussianities
and non-localities induced by non-linear gravitational evolution.

Another method of estimating covariances is Jack-knife resampling, which allows errors to be estimated internally, directly from the data \cite{Krewski81,Shao95}.
It does however require some arbitrary choices (number of Jack-knife regions, for example) and its performance is far from perfect 
(see e.g. Norberg et al. 2009). It also will not include fluctuations on the scale of the survey.  

Efforts to estimate covariance matrices directly from theory, that go
beyond a simple rescaling of the linear Gaussian covariance, must deal with non-linear evolution, shot-noise, redshift space distortions, and the complex mapping between galaxies and matter (Hamilton, Rimes and Scoccimarro 2006; Sefusatti et al. 2006; Pope and Szapudi 2008; de Putter 2012; Sefussati et al 2012). 

In this paper we present a new method for generating galaxy mocks that is significantly faster than mocks based on N-body simulations.  This method follows the main ideas put forward in the PTHalos method of Scoccimarro and Sheth (2002), but the implementation is overall simpler and differs significantly in some key aspects; the most relevant being that we do not use a merger tree to assign halos within big cells of the density field but instead we obtain the halos more precisely using a halo finder. This  method is fast because it is based on a matter field generated using 2nd Order Lagrangian Perturbation Theory (2LPT), but it still allows us to include the most important non-Gaussian corrections relevant for two-point statistics covariance matrices described by the trispectrum.


We use this method to create 600 mock galaxies catalogues occupying the volume required to accommodate the SDSS-III data release 9 (DR9) BOSS CMASS sample. This sample contains around a quarter million high-quality spectroscopic galaxy redshifts between $0.43 < z < 0.7$ distributed across an angular footprint over $3\,000$ sq. deg, and it represents the largest effective volume of any sample to date (see \citealt{boss12} for further details). We apply the CMASS DR9 selection function to the mock catalogues we create, thereby allowing the calculation of covariance matrices that include the full effect of the survey geometry. We thus provide the means by which statistical errors are determined for the CMASS DR9 sample.

This paper has two parts.  First, we describe our method for generating PTHalos and compare (and calibrate) it with N-body simulations from the LasDamas collaboration (McBride et al., in prep.).  In the second part, we populate the PTHalos with mock galaxies in a way that matches the CMASS sample. These mocks have been used in several analyses of BOSS DR9 data, including the study of systematics (Ross et al. 2012), the determination of the BAO scale \citep{boss12}, redshift space distortions (Reid et al. 2012, Samushia et al. 2012), evolution of galaxy bias (Tojeiro et al. 2012), the concordance with the $\Lambda$CDM model (Nuza et al. 2012), and the full shape of the correlation function (Sanchez et al. 2012). Note that the use of the mocks is not limited to only providing covariance matrices. For instance, by using mocks one can assess the level of expected chance correlation between galaxies and systematics (e.g. Ross et al. 2012).

Galaxy PTHalos mocks will be publicly available\footnote{http://www.marcmanera.net/mocks/}. 
A table with the monopole of the correlation function and the covariance matrix 
is given at the end of the paper. All log values in this paper are in base 10.

\section{Overview of the method}

Our goal is to develop a fast method for generating galaxy mocks, such that covariance matrices can be computed
accurately for galaxy samples such as the CMASS DR9 \citep{boss12} and the methods of analysis can be tested
for bias and relative accuracy. The basic steps in the method can be summarised as follows:
\begin{enumerate}
\item
Create a particle based 2LPT matter field (as described in Section \ref{sec:2LPT}).
\item
Identify halos using a Friends-of-Friends (FoF, Davis et al. 1985) halo-finder with an appropriately chosen linking length. We argue that, for the BOSS mean redshift, this linking length should be 0.38 times the comoving interparticle distance; see Section \ref{sec:PTH}.    
\item
Assign masses to halos by imposing a mass function that agrees with N-body simulations. 
\item
Populate halos with galaxies using a HOD algorithm calibrated to fit the observational data.  
\item
Apply the survey angular mask and galaxy redshift distribution. 
\end{enumerate}
We validate the first three steps by comparing our method with the clustering of halos in the N-body simulations whose halo abundances we have matched.  We then apply the final steps by calibrating the HOD to the CMASS DR9 dataset. Finally, we generate 600 mocks of CMASS galaxies with DR9 geometry and redshift selection.    

The gain in runtime achieved by generating PTHalos galaxy mock catalogues compared to creating mock catalogues from N-body simulations comes from the 
first step: for the particle numbers used here, 2LPT is about three orders of magnitude faster than N-body simulations. The time taken to make mock catalogues in PTHalos is dominated by the subsequent steps, and thus the speedup factor at the end of the procedure is reduced to about two orders of magnitude.

\section{Overview of the BOSS CMASS DR9 galaxies}

BOSS, part of the SDSS-III (Eisenstein et al. 2011) is an ongoing survey measuring spectroscopic
redshifts of 1.5 million galaxies, 160,000 quasars and a various
ancillary targets. BOSS uses SDSS CCD photometry (Gunn
et al. 1998,2006) from five passbands (\textit{u, g, r, i, z}; e.g.,
Fukugita et al. 1996) to select targets for spectroscopic observation.

The BOSS CMASS galaxy sample is selected with colour-magnitude cuts,
aiming to produce a roughly volume-limited sample in the redshift range of $0.4 < z < 0.7$, and results in a sample that is approximately stellar-mass limited. These galaxies have a bias of $\sim 2$ and most are central galaxies of halos of $10^{13} M_{\odot}$,
with a non-negligible fraction ($\sim 10$ per cent)being satellites in more massive halos
(White et al. 2011).

\begin{figure}
\center
\includegraphics[width=70mm]{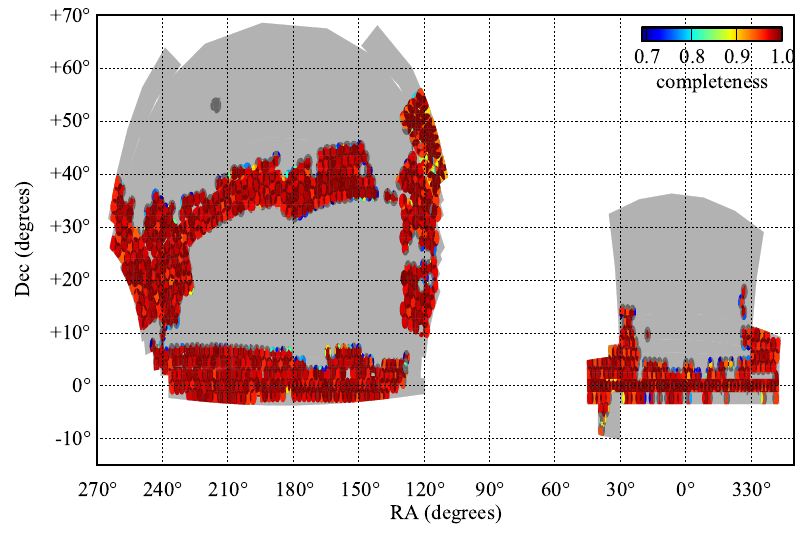}
\caption{The Northern Galactic cap (NGC) and Sourthern Galactic cap (SGC) footprint of the CMASS DR9 galaxy sample}
\label{bossmask}
\end{figure}

\begin{figure}
\center
\includegraphics[width=70mm]{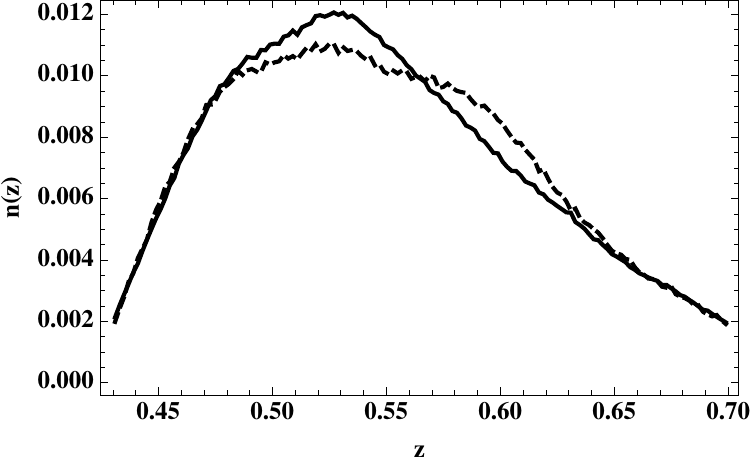}
\caption{Normalised redshift distribution of galaxies in the NGC (solid) and SGC (dashed) CMASS DR9 sample }
\label{nzfig}
\end{figure}

DR9 includes data taken up to the end of July 2011. The details of the catalogue and mask used for the large-scale
structure analyses are explained in \citet{boss12}, and an analysis of potential systematic effects is presented in Ross et al. (2012). 
DR9 covers approximately 3344 $\rm{deg}^2$ of sky (containing 264,283 usable redshift galaxies) 
of which 2635 $\rm{deg}^2$ (containing 207,246 galaxies) are in the Northern Galactic cap (NGC) 
and 709 $\rm{deg}^2$ (containing 57,037 galaxies) are in the Southern Galactic cap (SGC), as
shown in Figure \ref{bossmask}. The NGC and SGC have slightly different redshift distribution of
galaxies; we show their normalised redshift distributions, $n(z)$, in
Figure \ref{nzfig}. NGC and SGC mock catalogues have been generated
according to these distributions.

\section{Summary of 2LPT}
\label{sec:2LPT}

\subsection*{Basics of Lagrangian Perturbation Theory} 
\label{sub:2lptbasics}

The Lagrangian description of structure formation \citep{Buc89,Mou91,Hiv95}
relates the current (or Eulerian) position of a mass element, $\x$, to
its initial (or Lagrangian) position, $\mathbf{q}$, through a displacement
vector field $\mathbf{\Psi}(\mathbf{q})$,
\begin{equation}
\x = \q + {\bf \Psi}(\q).
\label{xpsi}
\end{equation}
The displacements can be related to overdensities by (Taylor and Hamilton 1996)
\begin{equation}
 \delta(\mathbf{k}) = \int d^3q\ e^{-i \mathbf{k}\cdot q}
 \left(e^{-i \mathbf{k}\cdot \mathbf{\Psi}(\mathbf{q})} - 1\right) .
\label{eqn:lptdensity}
\end{equation}

Analogous to Eulerian perturbation theory, LPT expands the displacement in
powers of the linear density field, $\delta_L$,
\begin{equation}
 \mathbf{\Psi} = \mathbf{\Psi}^{(1)} + \mathbf{\Psi}^{(2)} + \cdots \; ,
\label{eqn:psiexp}
\end{equation}
with $\mathbf{\Psi}^{(n)}$ being $n^{\rm th}$ order in $\delta_L$.
First order in LPT is equivalent to the well-known Zel'dovich approximation (ZA).

The equation of motion for particle trajectories $\x(\tau)$ is

\beq
\frac{d^2 \x}{d \tau^2} + {\cal H}(\tau) \ \frac{d \x}{d \tau}= -
\del \Phi,
\label{motion}
\eeq
where $\del$ the gradient operator in Eulerian coordinates $\x$ and $\tau$
is conformal time. Here $\Phi$ denotes the gravitational potential,
and ${\cal H}=\frac{d ln a}{d\tau}=Ha$ denotes the conformal expansion 
rate. $H$ is the Hubble factor and $a$ the scale factor.

Substituting (\ref{xpsi}) into (\ref{motion}) and solving the equation at
linear order gives the Zel'dovich (1970) approximation (ZA),

\beq
\del_q \cdot {\bf \Psi}^{(1)}= -D_1(\tau) \ \d(\q),
\label{Psi1}
\eeq

\noindent where we have taken a gradient of Eq \ref{motion}
and used the Poisson equation to relate $\Phi$ and $\d(q)$.
Here $\d(q)$ denotes the Gaussian density field imposed
by the initial conditions and $D_1(\tau)$ is the linear growth
factor. In Eq \ref{Psi1} the gradient is in Lagrangian coordinates $q$,
while in Eq \ref{motion} it is in Eulerian coordinates; the two are
related by the Jacobian of the coordinate transformation.

The solution to second order describes the correction to the
ZA displacement due to gravitational tidal effects and reads

\beq
\del_q \cdot {\bf \Psi}^{(2)}= \frac{1}{2} D_2(\tau) \sum_{i \neq j}
(\Psi_{i,i}^{(1)} \Psi_{j,j}^{(1)} - \Psi_{i,j}^{(1)} \Psi_{j,i}^{(1)}).
\label{Psi2}
\eeq

\noindent where the comma followed by a coordinate denotes partial derivative
in that direction.
 
Since Lagrangian solutions up to second-order are curl-free, 
it is convenient to define two Lagrangian potentials
$\phi^{(1)}$ and $\phi^{(2)}$ (${\bf \Psi^{(i)}}=\del_q\phi^{(i)}$) 
so that the solution up to second order reads 
\beq
\x(\q) = \q -D_1\ \del_q \phi^{(1)} + D_2\ \del_q \phi^{(2)}.
\label{dis2}
\eeq
\noindent Likewise one can solve for the velocity field, which reads 
\beq
{\bf v} = -D_1\ f_1\ H\ \del_q \phi^{(1)} + D_2\ f_2\ H\ \del_q \phi^{(2)}.
\label{vel2}
\eeq
Here $v \equiv \frac{d x}{d t}$ is the peculiar velocity, 
$t$ denotes cosmic time, $f_i=\frac{dln D_i}{dln a}$, and 
$D_2$ denotes the second order growth factor.  
To better than $0.6$ per cent accuracy, 
\begin{equation}
D_2(\tau) \approx -\frac{3}{7} D_1^2(\tau) \Omega_m^{-1/143},
\end{equation}
for values of $\Omega_\Lambda$ between 0.01 and 1 (Bouchet et al. 1995). Lagrangian perturbation theory has been used to model baryon acoustic oscillations (Matsubara 2008a,b; Padmanabhan \& White 2009; Padmanabhan, White \& Cohn 2009).
For a more detailed explanation of 2LPT see Bernardeau et al. (2002) and references therein.

To generate the 2LPT displacement we used an algorithm that takes advantage of fast Fourier transforms (FFT) and is described in detail in Scoccimarro (1998). Although this algorithm assumes Gaussian initial conditions, it can be extended to treat non-Gaussian initial conditions given by any factorisable primordial bispectrum (Scoccimarro et al. 2012), and a parallel version of such code is publicly available\footnote{http://cosmo.nyu.edu/roman/2LPT/ and http://www.marcmanera.net/2LPT/}. In this paper we only consider Gaussian initial conditions, although the same procedure we describe can be applied to the primordial non-Gaussian case.

Compared to Scoccimarro \& Sheth (2002) our implementation of 2LPT differs only in the smoothing applied to the linear density field before constructing the Zel'dovich displacement field. To reduce the effects of orbit crossing (where Lagrangian perturbation theory breaks down), they impose a cutoff in the linear spectrum, similar to the standard truncated Zel'dovich approximation (Coles et al. 1993).  
However, rather than using their merger tree method to identify halos, here we identify halos by applying the FoF algorithm to the 2LPT field with a modified linking length. The theoretical motivation for the choice of linking length can be derived from the spherical collapse in 2LPT dynamics (see Section~\ref{sec:linktheo}). In order to preserve this theoretical choice, we would like 
to change the linear density field on smoothing scales of the order of the Lagrangian size of halos as little as possible, while at the same time not have excessive orbit crossing effects for the halos that host the galaxies we are interested in. These competing requirements become increasingly difficult to satisfy as the halo mass we are interested in decreases. Although we have not done an exhaustive investigation, a smoothing window described the linear density field Fourier amplitudes multiplied by ${\rm e}^{-k /(4 + k)/2}$ (with $k$ in h/Mpc) works reasonably well for the halo mass range relevant for our purposes; see Section \ref{sec:PTH}. On top of this, there is of course a sharp-cutoff in the linear spectrum at the Nyquist frequency of the particle grid used to generate the fields (with grid size $N_{\rm grid}=1280$).

\section{Cosmology and resolution specifications}
\label{sec:lasdamas}

We have produced halo and galaxy mocks using two different sets of 
$\Lambda$CDM cosmological parameters. The first set has been 
chosen to match that of the N-body simulations we use  
to calibrate the PTHalos method, while the second set has
been chosen to have values closer to those expected from observations.  
\\
\\
{\textit{LasDamas Cosmology}:} 

The fiducial parameters for this cosmology are: 
$\Omega_m = 0.25$, $\Omega_\Lambda=0.75$, $\Omega_b = 0.04$, $h=0.7$,
$\sigma_8=0.8$ and $n_s=1$. 
These parameters were used by the LasDamas collaboration\footnote{http://lss.phy.vanderbilt.edu/lasdamas/} which produced a 
suite of large N-body cosmological dark matter simulations (McBride et al., in prep).  
These simulations were run with a TreePM code Gadget-II (Springel 2005), with a FFT grid size of 2400 points in each dimension. 
Each simulation run covers a cubical volume of a box size L=2400 Mpc/h, and have 1280$^3$ dark matter particles. We have created PTHalos mocks assuming the same cosmology and resolution parameters, so as to properly compare halo clustering in each of the 40 N-body simulation runs, and thus calibrate our method.  As shown in Section \ref{sec:PTH}, we achieve a 10 per cent accuracy in the clustering of halos.
\\
\\
{\textit{WMAP Cosmology}:} 

The second $\Lambda$CDM cosmology that we consider has the following parameters:
$\Omega_m = 0.274$, $\Omega_\Lambda=0.726$, $\Omega_b = 0.04$, $h=0.7$,
$\sigma_8=0.8$ and $n_s=0.95$.
These are the same as those used to analyse the first semester of BOSS data (White et al. 2011) and in the \citet{boss12} analysis; they are within $1\sigma$ of the best fit WMAP7 concordance cosmological model \citep{larsonwmap}.

We have two simulations of $3000^3$ particles and cubical box size of L=2750 Mpc/h with which we compare the clustering of our runs. These simulations were performed with 
the TreePM code described in White (2010), which  has been compared to a number of other codes and shown to achieve the same precision level as other N-body codes for such simulations \citep{Hei08}. We use one of these simulations 
in Section \ref{sec:wmapsim}.
\\
\\
{\textit{Resolution Parameters}:} 

We run 2LPT for our mocks in a cubical box of of size $L=2400$ Mpc/h with $N=1280^3$ particles.  This matches the specifications of the Oriana simulations of LasDamas suite, and allows us to easily match the Fourier phases in 2LPT runs to those of the Oriana simulations, thus allowing a direct comparison for each realisation.  With these parameters the mass resolution for the LasDamas and WMAP cosmologies is $M_{part}=45.7\cdot 10^{10} M_{\odot}/h$ and $50.1\cdot 10^{10} M_{\odot}/h$, respectively.  The cubical box was matched to the CMASS DR9 geometry as explained in Section \ref{sec:geometry}.


\section{PTHalos} 
\label{sec:PTH}
The first step is to generate a 2LPT field, as described in Section \ref{sec:2LPT}, which is traced by means of a distribution of particles. Based on this field,
halo positions and raw masses are found by means of a FoF algorithm, which percolates all pairs of particles separated by a distance $d \leq b$. This algorithm has become a standard technique and has been used extensively in astrophysics and cosmology since Davis et al. (1985). 
Using the LasDamas simulations we calibrated the FoF linking length, and set $b=0.38$ times the mean interparticle separation as the value for generating mocks.
Note that this is substantially larger than the usual choice, $b=0.2$, in N-body simulations.  Section \ref{sec:linktheo} shows that this choice is motivated by 2LPT dynamics.

The second step of the method is a reassignment of halo masses. Respecting the ordering given by the FoF number of particles, 2LPT halo masses are changed so that the mean mass function of PTHalos matches a given fiducial mass function. The underlying understanding here is that the ranking of the masses is more accurate than their exact values, which will vary according to the definition of halo boundaries, both in N-body simulations and 2LPT runs.

Note that, given a fiducial mass function for PTHalos, a fixed 2LPT halo mass always corresponds to the same PTHalo mass. That is, the mapping of the masses is between the \emph{mean} of 2LPT realizations of the mass function and the targeted fiducial one. In this way, the scatter of the measured mass function between 2LPT realisations is translated, as expected, into a scatter of the PTHalos mass function. 

In this paper, the PTHalos realisations with the LasDamas cosmology have as a given mass function that of the mean
of the LasDamas N-body simulations. The PTHalos realisations with WMAP cosmology use the mass function
of Tinker et al. (2008), and adopt the definition of dark matter halos that correspond to overdensities $200$
times the mean background density.

\subsection{Linking length: Theoretical motivation}
\label{sec:linktheo}

\begin{figure}
\center
\includegraphics[width=80mm]{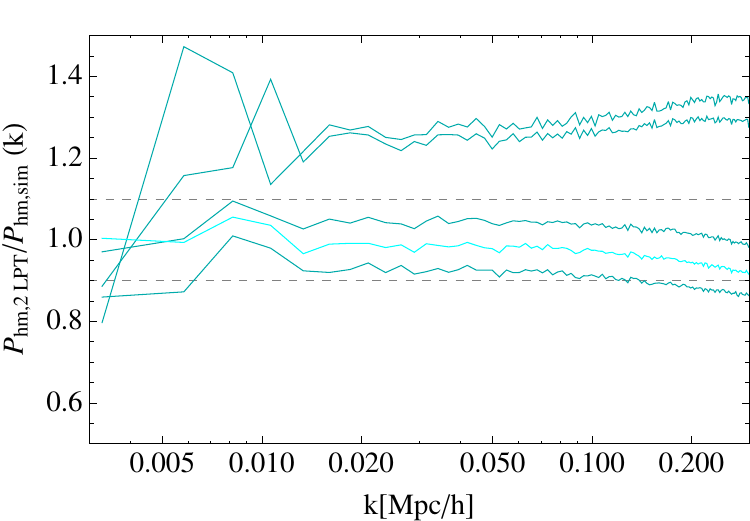} 
\caption{Ratio between PTHalos and N-body halo-matter cross-power spectra as a function of linking length, $b$, for the $10^6$ most massive halos. From top to bottom linking length are: 0.27, 0.30, 0.36,0.38 (in lighter color), and  0.40. N-body halos use $b=0.2$ with the corresponding mass threshold of $3.02\cdot 10^{13} M_{\odot}/h$} 
\label{crosspowerlogn6}
\end{figure}

The appropriate FoF linking length in Eulerian N-body simulations is roughly determined as follows.  Given $\Omega_m$ and $\Omega_\Lambda$ one uses a fitting function (see Eq. \ref{BryanDvir}) to compute the virial overdensity $\Delta_{vir}$ of halos within the spherical infall model.  For the LasDamas cosmology, 
$\Delta_{vir}=377$ times the mean background density, at redshift zero.

Then, assuming an isothermal profile for the dark matter halo, one can relate the mean density of the halo to the density at the virial radius, i.e, $\rho_{Rvir}= \Delta_{vir}/3$. This density is converted to a mean separation of particles by assuming that the density at the virial radius is equal to that of two particles in a sphere of radius $b$.  
For the LasDamas cosmology, this gives $b=0.156$ in units of the mean interparticle separation. For an Einstein de Sitter cosmology, $\Omega_m=1$, $b=0.2$, which is the value most commonly used in the literature.

Now, because the 2LPT dynamics is an approximation to the
true dynamics of the dark matter field, it yields halo densities
that consistently differ from the N-body densities. Consequently, the FoF linking 
length of 2LPT matter field, $b_{2LPT}$, needs to be rescaled from the value used in N-body
simulations, $b_{sim}$. The rescaling is given by

\begin{equation}
b_{2LPT} = b_{sim} \left( \frac{\Delta_{vir}^{sim}}{\Delta_{vir}^{2LPT}} \right)^{(1/3)} \; .
\label{linkscaled}
\end{equation}

Both the halo virial overdensity in N-body simulations, $\Delta_{vir}^{sim}$, and its corresponding value
in the 2LPT field, $\Delta_{vir}^{sim}$, are easy to compute. For the N-body case we take the
value of Bryan and Norman (1998),
\begin{equation}
\Delta_{vir}^{sim}=(18\pi^2+82(\Omega_m(z)-1)-39(\Omega_m(z)-1)^2) /\Omega_m(z)\; ,
\label{BryanDvir}
\end{equation}
where
\begin{equation}
\Omega_m(z)=\Omega_m (1+z)^3/H(z),
\end{equation}
which gives $\Delta_{vir}=244$ at redshift z=0.52. We chose this redshift
because it is the redshift at which we will compare with LasDamas simulation 
outputs, and it is close to the mean redshift of the BOSS CMASS sample, 
for which we want to produce galaxy mock catalogues. 

The Lagrangian $\Delta_{vir}^{2LPT}$ can be easily obtained from the relation between Lagrangian and Eulerian fields, 
which are related by the determinant (Jacobian) of the transformation of Eq. (\ref{xpsi}),

\begin{equation} 
\delta_{LPT} = \left( {\rm{Det}}(1+ \partial \Psi_i/\partial q_j) \right)^{-1} -1\; .
\end{equation}

Having solved Equations (\ref{Psi1}) and (\ref{Psi2}), and thus knowing $\Psi$ at second order, this equation can be rewritten in terms of the growth factor. Then, assuming spherical symmetry for simplicity, it reads

\begin{equation} 
\delta_{2LPT} = \left( 1-\frac{\delta_0}{3}(D_1-\frac{\delta_0}{3}D_2) \right)^{-3} -1\; ,
\end{equation}

\noindent where $\delta_0$ is the over-density at the initial time. Since we know from spherical collapse in Eulerian dynamics that a halo has virialised when its linear density fluctuation is $D_1\delta_0=1.686$, we can predict the 2LPT over-density of at virialization to be
$\Delta_{vir}^{2LPT}=\delta_{vir}^{2LPT}+1=35.43$ times the mean background density.

Therefore, using Equation (\ref{linkscaled}) we find that to conduct a robust comparison between PTHalos halos and N-body halos of linking length of $b=0.2$, we need to use a linking length of $b=0.38$ in the 2LPT field. It is worth emphasising that this predicted value is approximate. The actual value has to be confirmed (or otherwise set) by comparing the clustering of halos between 2LPT and N-body simulations. This processs is described in the next Section, where we find that this value works very well at the 10 percent level.

In principle, one can use spherical overdensity (SO) methods to identify halos instead of the FoF algorithm \citep{Lacey94}. A similar procedure to that discussed in Section \ref{sec:linktheo} could then be used to match the SO density parameter in N-body simulations to 2LPT.

\subsection{Linking length: Calibration with N-body simulations}
\label{sec:bcal}

\begin{figure}
\center
\includegraphics[width=80mm]{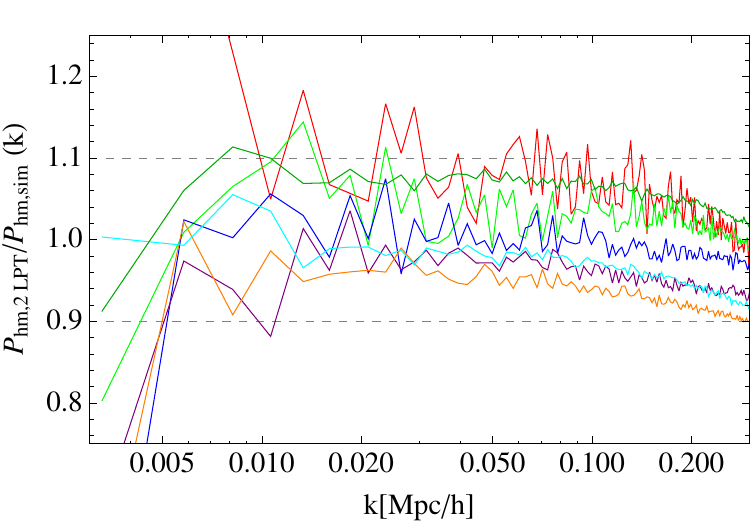} 
\caption{Ratios between PTHalos and N-body halo-matter cross-power spectra as a function of halo mass threshold, for a linking length of b=0.38 (2LPT) and b=0.2 (N-body). The halo masses are given in Table \ref{masstable}.} 
\label{crosspowermany}
\end{figure}

\begin{table}
\begin{center}
\begin{tabular}{ccl}
$\log N$ & Mass ($N_{\rm part}$) & Color \\
\hline
3.5  & 44.3 (968)  &  Red \\
4.0  & 30.7 (672)  &  Light Green\\
4.5  & 19.8 (432)  &  Blue \\
5.0  & 11.7 (256)  &  Purple \\
5.5  & 6.31 (138)  &  Orange \\
6.0  & 3.02 (66)   &  Cyan \\
6.5  & 1.28 (46)   &  Dark Green\\
\end{tabular}
\end{center}
\caption{ The number of halos, their mass, and associated color. Masses of halos in N-body simulations as a function of their position in the mass-ranked list. That is, given the $N$ most massive halos in the volume $L=(2400 Mpc/h)^3$ the lower mass in the sample is $M$. Masses are from run 1001 of Oriana N-body Simulation at $z=0.52$ and are given for the linking length of $b=0.2$. Masses are in units of $10^{13} M_{\odot}/h$ and are not corrected for discreteness effects. For each halo mass we have shown in parentheses the number of particles that that halo has given our mass resolution.}
\label{masstable}
\end{table}

In order to confirm the linking length that we need to use for PThalos,
we have run a 2LPT simulation at z=0.52 with the same Fourier phases and amplitudes
as that of one of the Oriana simulations from the LasDamas collaboration.

    We obtained halos from the 2LPT dark matter field using FoF with different linking lengths around the value given by Eq \ref{linkscaled}  We then computed the cross-power spectrum between the 2LPT halos and the N-body matter field, $P_{hm,2lpt}(k)$, and the cross-power spectrum between the N-body halos and N-body matter field, $P_{hm,sim}(k)$, where these latter halos, from LasDamas, where obtained with $b=0.2$.

    The comparison between these two spectra gives a measure of accuracy of the bias of the 2LPT halos. In particular we are interested in comparing the ratio  $P_{hm,2lpt}/P_{hm,sim}$ since this is equivalent to the ratio of the halo bias factors.  Note that we have computed the cross-power spectra and not the auto-power spectra since in this case we do not need to correct our estimator for shot-noise. 

    The results are shown in Figures \ref{crosspowerlogn6} and \ref{crosspowermany}.
Figure \ref{crosspowerlogn6} shows the ratio $P_{hm,2lpt}/P_{hm,sim}$ of the 
million most massive halos as a function of the wavenumber $k$
for different values of the linking length.
We see that, as we increase the linking length, the ratio of the cross-powers decreases. There
is a range of linking lengths ( $0.36 < b < 0.40$ ) for which the ratio of the
bias is within 10 per cent. $b=0.38$, as computed using Eq \ref{linkscaled} performs best.

    The sample of halos of Figure \ref{crosspowerlogn6} is equivalent to a mass threshold of
$M=3.02\cdot 10^{13} M_{\odot}/h$. We have found that $b=0.38$ also fits the other halo
samples with different mass thresholds within 10 per cent accuracy. In fact, it is the
linking length that performed best across the range of halo masses we consider, and that are shown in Table \ref{masstable}.
The ratios $P_{hm,2lpt}/P_{hm,sim}$ for these halos are plotted in Figure \ref{crosspowermany},
with the colours referenced in Table \ref{masstable}. All the curves are within 10 per cent accuracy,
which justifies our choice to set $b=0.38$, in accordance with the theoretical expectation, as our fiducial value for the 2LPT linking length.

\subsection{Variance and Cross-Correlation Coefficients}

\begin{figure}
\center
\includegraphics[width=80mm]{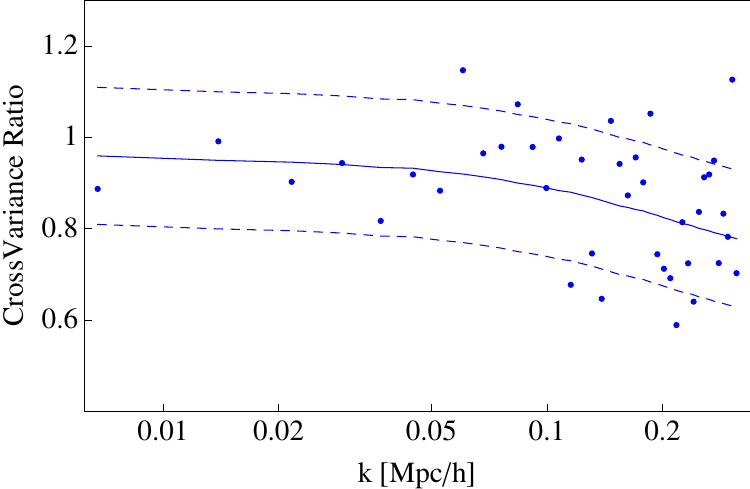}
\includegraphics[width=80mm]{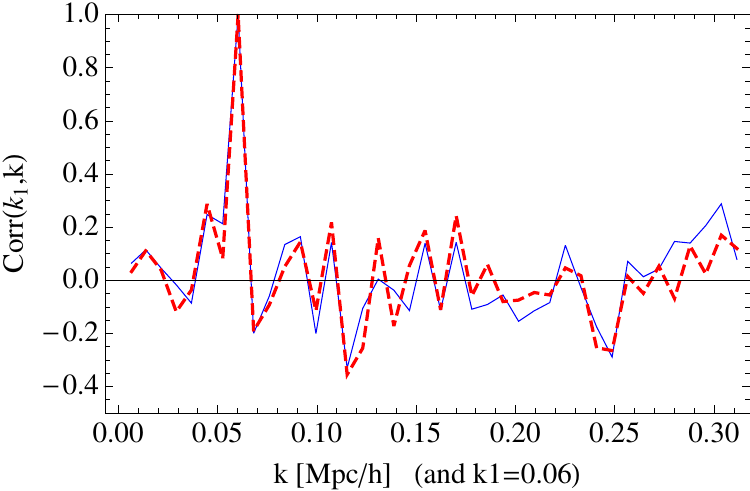}
\includegraphics[width=80mm]{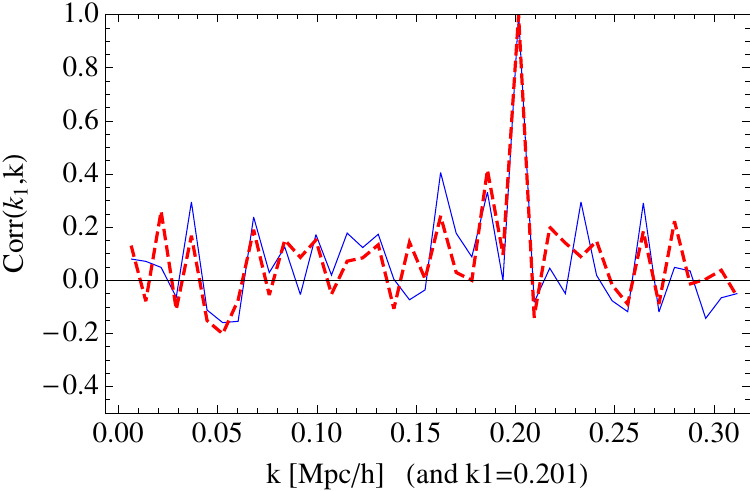}
\caption{TOP: Ratio of the cross-power variance of PTHalos and N-body simulations for a mass threshold of $3.02\cdot10^{13}M_{\odot}/h$.
The expected ratio is shown in a solid line and a 15 per cent band range is shown in dashed lines.
MIDDLE and BOTTOM: Comparison of correlation coefficients of N-body (solid blue) and PTHalos (dashed red) halo-matter cross-power spectra.
Middle panel $k_1=0.06$. Bottom $k_1=0.201$}
\label{covariance40}
\end{figure}

Having set the PThalos linking length, we run 40 2LPT dark matter fields, with the same Fourier phases and amplitudes as the 40 simulations of LasDamas suite. For all of them we compute the halo-matter cross-power spectra as in the previous Section, for the first million halos of both the PTHalos and N-body halos, and we compute the corresponding covariance matrices:

\begin{eqnarray}
&&C(k_i,k_j) = \\
&& \frac{1}{N-1}\sum_{i=1}^{N=40}  (P_{hm}(k_1) -\bar{P}_{hm}(k_1))  (P_{hm}(k_2)-\bar{P}_{hm}(k_2)) \; ,
\nonumber
\end{eqnarray}
where $\bar{P}_{hm}$ is the mean power spectrum of the set of simulations. The 
variance of the cross-power spectrum is defined as ${\rm CrossVariance}={\rm Var}(k)=C(k,k)$
and the correlation coefficients of a given $k_1$ as ${\rm Corr}(k_1,k)=C(k_1,k)/\sqrt{{\rm Var}(k_1) {\rm Var} (k)}$.

In the top panel of  Figure \ref{covariance40} we show the ratio of the variances between PTHalos and N-body simulations,
for the first million halos, which is an equivalent mass threshold of $3.02\cdot10^{13}M_{\odot}/h$.
We can see that most points lie within 15 per cent range of the expected value at linear order, which is
given by the square of the ratio of the halo bias. Since the bias of the halos is accurate at 10 per cent
the variance is accurate at about 20 per cent. 

In the middle and bottom panels of Figure \ref{covariance40} we show a comparison between the correlation coefficients  
of PTHalos (dashed red) and N-body  simulations (solid blue). Each line shows the mean of the 40 realizations
that have the same phases. Both are clearly similar, showing that 
the PTHalos preserve the same structures as the N-body simulations.

\subsection{Autocorrelation}

\begin{figure} 
\center
\includegraphics[width=70mm]{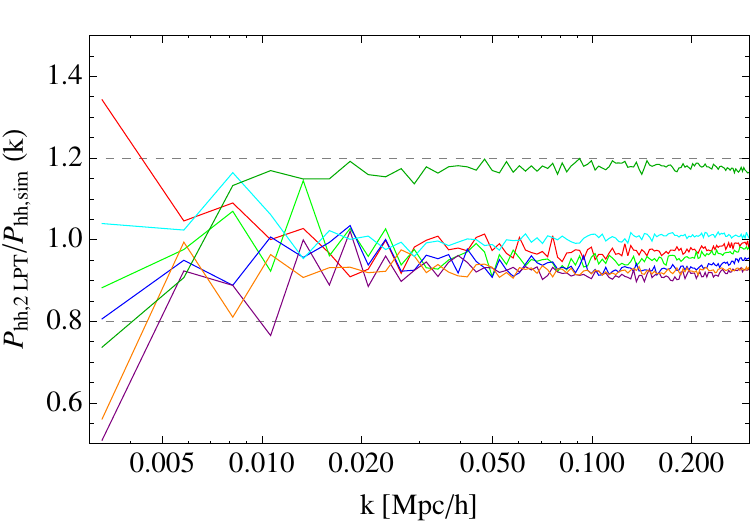}
\caption{Ratios between 2LPT and N-body halo power spectra as a function of $k$ for different halo mass thresholds, for a linking length 
    of b=0.38 (2LPT) and b=0.2 (N-body). The correspondence between color and halo mass thresholds are given in Table~\ref{masstable}. The power spectra have not been corrected for shot-noise.} 
\label{crosspower5}
\end{figure}

\begin{figure}
\center
\includegraphics[width=70mm]{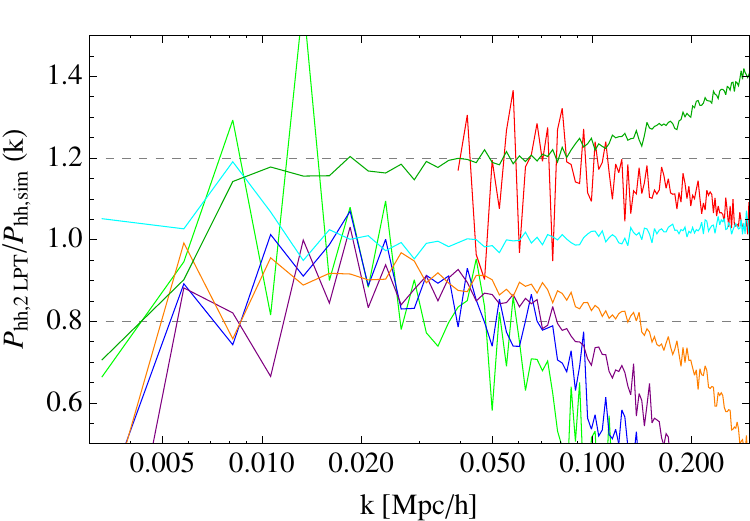}
\caption{Same as Fig.~\ref{crosspower5} but with shot-noise corrected power spectra, assuming Poisson noise.}  
\label{crosspower4}
\end{figure}

In Figure \ref{crosspower5}, we show the ratio between the auto-power spectrum of PTHalos and the 
auto-power of N-body halos, where we did not subtract a shot noise contribution. We see that for 
most of the masses this ratio is within 10 per cent, except for the
halos of our lower mass range (dark green line) that are significantly more clustered than the 
N-body halos of equivalent mass. This is likely due to a fraction of small 
halos being clustered around massive ones, probably because of the shell crossing effects that 
makes halos in 2LPT less compact than in N-body simulations.  

In Figure \ref{crosspower4} we show the ratio of the auto-power spectra for 
one realization of PTHalos and an N-body simulation from LasDamas with the same Fourier phases. 
Before doing the ratio, a Poisson shot noise contribution of $1/n$ 
(where $n$ is the number density of halos) was subtracted from the power, 
as is common under the approximation of Poisson sampling. Note however, 
that there are indications in the literature 
that the shot-noise of halos is not strictly Poisson (Appendix A, Smith et al. 2007).  
As seen in Figure \ref{crosspower4}  we recover a ratio within $\sim 20$ per cent for most masses
and range of scales, which is consistent with our findings of an accuracy of 10 per cent in 
the ratio of the bias (or equivalently of the cross-power spectra). At small-scales, for $k > 0.15\, h/$ Mpc, PTHalos performance decreases
significantly. This is the region between the one halo and the two halo terms of the
correlation function, and corresponds to the scales that are more difficult to model with 
our simple method.

\subsection{PTHalos with WMAP cosmology}
So far we have established a method to obtain halos from a 2LPT dark 
matter field, which matches the clustering of simulations at 10 per cent. We have tested the 
method by comparing the clustering of PTHalos with that from the LasDamas N-body simulations suite. 

In the rest of this paper, we use our WMAP fiducial cosmology, 
which is closer to that expected from observations.  

Using our PTHalos code we have generated 600 2LPT fields at z=0.55. 
PThalos were obtained using a linking length of $b=0.38$. For these runs,
since we cannot use LasDamas mass function to set the mass of the halos 
and instead use a general description of Tinker et al. (2008), using SO halos
corresponding to 200 times the mean background density. 

\begin{figure}
\center
\includegraphics[width=70mm]{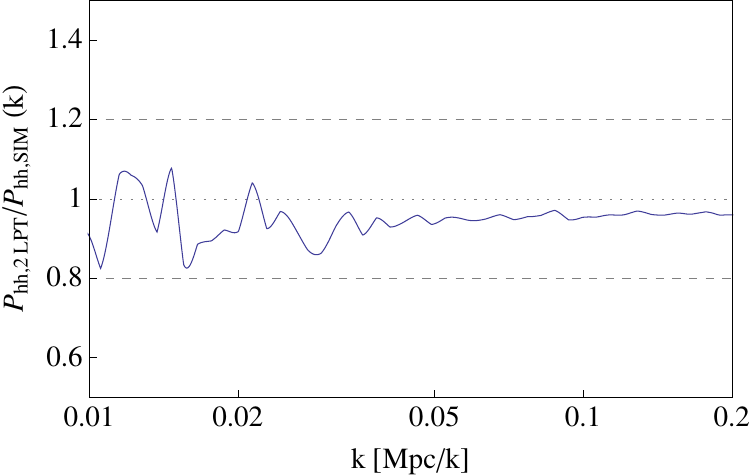}
\caption{The ratio between the average power spectrum calculated from
  PTHalo simulations and the power spectrum calculated for halos
  selected from the White et al. (2011) simulation.  For both we apply
  a mass cut of $10^{13} M_{\odot}/h$. Poisson shot-noise ($1/n$) has
  been subtracted.}
\label{MWhalos}
\end{figure}

We do not expect the change in cosmological model to significantly affect the 
accuracy of the PTHalos method. Nonetheless, we have compared the PTHalos
clustering with the clustering of the N-body simulation of White et al. (2011) for halos above $10^{13} M_{\odot}/h$. This N-body simulation  
reproduces a piece of the universe with the same cosmological parameters that we use in the
remaining of the paper. Their halos are identified with a FoF algorithm with b=0.168, but the 
clustering is still matched at the 10 per cent level. This can be seen in Figure \ref{MWhalos}
where we have plotted the ratio of the halo power spectrum calculated from the N-body 
simulation and the PThalos method. This result shows the robustness of the PTHalos method.

\begin{figure}
\center
\includegraphics[width=70mm]{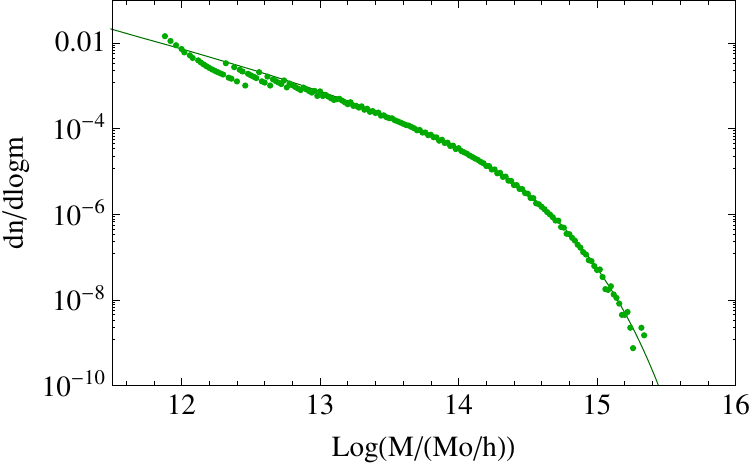}
\caption{Comparison of mass functions from the simulation of White et
  at 2011, assuming a friends-of-friends parameter $b=0.168$, and the
  Tinker et al. (2008) mass-function fitting function for halos corresponding 
  to 200 times the mean background density.}
\label{MWmass}
\end{figure}

For this N-body simulation we also show in Figure \ref{MWmass} the mass function of the
halos together with that of Tinker et al. 2008, which is the
mass function we used to set the masses of PThalos for our fiducial WMAP cosmology.  
As expected the fit is good except at the low mass end where the mass resolution effects of 
our simulation start to become important.  

\label{sec:wmapsim}

\section{Populating halos with galaxies}

\subsection{Halo Occupation Distribution}

To populate halos with galaxies we use a Halo Occupation Distribution (HOD; Peacock \& Smith 2000, Scoccimarro et al. 2001, Berlind \& Weinberg 2002)
 functional form with five parameters, as used by Zheng et al. (2007). In this form, the mean number of galaxies in a halo of mass
$M$ is the sum of the mean number of central galaxies plus the mean number of 
satellite galaxies, $N(M)= \langle N_{cen}(M)\rangle + \langle N_{sat}(M)\rangle $,where 

\begin{eqnarray}
\langle N_{cen} \rangle  &=& \frac{1}{2} \left[ 1 + {\rm{erf}}\left( \frac{{\rm log} M - {\rm log} M_{min} }{\sigma_{{\rm log} M}}\right)\right] 
\nonumber \\
\langle N_{sat}\rangle &=& \langle N_{cen}\rangle \left( \frac{M - M_0 }{M_1}\right)^\alpha \; , 
\label{hod}
\end{eqnarray}

\noindent and $N_{sat}=0$ if the halo has $M < M_0$. The error function characterises the
scatter between the mass and the luminosity of the central galaxy, and the power law in the satellite 
occupation term characterises the efficiency of galaxy formation on mass.
The exact value of the HOD parameters that we use was determined by
fitting the DR9 galaxy clustering data, as explained in Section \ref{sec:fit}. 
The existence or not of a central galaxy in each halo is given by the $N_{cen}$ probability, and the
number of satellites will be drawn from a Poisson distribution with mean value $N_{sat}$.
In the rare event that we draw one satellite galaxy but no central one, we 
treat it as a central.  

\subsection{Halo profile} 

We have distributed satellite galaxies within a halo following 
a NFW density profile \citep{NFW}:

\begin{equation}
\rho(r) = \frac{4 \rho_s}{\frac{r}{r_s}\left(1+\frac{r}{r_s}\right)^2} \; , 
\end{equation}

\noindent where $r_s$ is the characteristic radius where the profile has a slope of $-2$,
and $\rho_s$ is the density at this radius. The ratio between the virial radius
$R_{vir}$ and the characteristic radius gives the concentration parameter,

\begin{equation}
c = \frac{R_{vir}}{r_s}.
\end{equation}

The masses of the halos and their concentrations are related. For our galaxy mocks
we use the relation found by Prada et al. (2011) when fitting data from N-body simulations.

\begin{eqnarray}
  c(M,z)  &=& \frac{B_0(x)}{B_0(1.393)},\, \mathcal{C}(\sigma')  \\
  \sigma'&=& \frac{B_1(x)}{B_1(1.393)}\, \sigma(M,x)  \nonumber \\
  \mathcal{C}(\sigma') &=& A \, \left[\left(\frac{\sigma'}{b}\right)^c+1\right] 
                  \exp\left(\frac{d}{\sigma'^2}\right), \nonumber 
\end{eqnarray}
where
\begin{eqnarray}
B_0(x) & = & c_0+ (c_1- c_0) 
    \left[\frac{1}{\pi}\arctan\left[\alpha(x-x_0)\right] + \frac{1}{2} \right] \nonumber  \\
B_1(x) & = & \frac{1}{\sigma_0}+ (\frac{1}{\sigma_1}-\frac{1}{\sigma_0}) 
     \left[\frac{1}{\pi}\arctan\left[\beta(x-x_1)\right] +\frac{1}{2} \right], \nonumber \\
\end{eqnarray}

\noindent and the parameters from the N-body fit are: $A=2.881$, $b=1.257$, $c=1.022$, $d=0.060$,  
$c_0 = 3.681$, $c_1 = 5.033$, $\alpha = 6.948$, $x_0 = 0.424$,
$\sigma^{-1}_0 = 1.047$, $\sigma^{-1}_1 = 1.646$, $\beta = 7.386$, $ x_1 = 0.526$.

The cosmology and redshift dependence of the fit enters through $x=(\Omega_\Lambda/\Omega_m)^{1/3}/(1+z)$ 
and through the variance of the halos of a given mass, $\sigma(M,z)$. The masses in the equations above
are defined such that the mean density at the virial radius is $200$ times the critical density.
Since we are using the Tinker et al. (2008) definition to set the masses, our PTHalos are closer to $200$ 
times the background density. Using the NFW we can easily move from one definition of halo mass to another,
and use each formula appropriately. 

We have added a dispersion to the mass-concentration
relation. We use a lognormal distribution, thus the probability of a 
concentration $c$ for a halo of mass M is, 

\begin{equation}
p(c|M) = \frac{1}{c \sqrt{2\pi\sigma_{log c}^2}} 
{\rm{Exp}} \left[ -\frac{ { {\rm{log}}[c/\bar{c}(M,z)] }^2 }{2\sigma_{log c}^2}\right]
\end{equation}
where $\bar{c}$ is the mean mass-concentration relation. Typical values of $\sigma_{logc}$ are
between $0.043$ and $0.109$ (Giocoli et al. 2010). We have chosen for our mocks  
the value $\sigma_{logc} = 0.078$, which is close to the mean. 

The scatter of the mass-concentration is not dependent on cosmological parameters
\citep{Maccio08}.

\subsection{Galaxy velocities in halos}

We assign velocities to the galaxies in halos by using the virial theorem, which states 
that the average kinetic energy of particles is half the average of the negative potential energy,
$\langle v^2\rangle = \langle G M(r)/r \rangle$. This 
average over the dark matter particles can be expressed as an integral of dark matter profile,

\begin{equation}
\langle v^2 \rangle =  \frac{\int_0^R G M(r) dM}{\int_0^{R_{vir}} dM}. 
\end{equation} 

Assuming a NFW profile, the mass inside a given radius is,

\begin{equation}
M(r) = \frac{4 \pi}{3}\rho_s^3 \cdot 12\left[ \ln(1+x) - \frac{x}{1+x}\right], \nonumber  
\end{equation} 

\noindent and therefore, the virial velocity reads

\begin{eqnarray}
\label{eqFc}
\langle v^2 \rangle & = & \frac{ G M }{R_{vir}}\cdot c \frac{\frac{c}{2(1+c)}-\frac{ln(1+c)}{1+c} }{\left(ln(1+c)-\frac{c}{1+c}\right)^2} \\ \nonumber 
       = \frac{ G M }{R_{vir}} & c &\frac{ 0.5\, c (1+c) - (1+c) ln(1+c) }{\left((1+c)ln(1+c)-c\right)^2} \\ \nonumber
       \equiv \frac{ G M }{R_{vir}} & F(c)  & 
\end{eqnarray} 

\noindent where the last equality defines $F(c)$ that we will use later. 
Here, again, $c$ denotes the concentration parameter, $c=R_{vir}/r_s$, $r_s$ denotes 
the characteristic NFW radius, and $R_{vir}$ denotes
the virial radius, defined as the radius at which the average density of 
the halo is $\Delta$ times the mean density $\bar{\rho}$, $M=4\pi /3 R^3 \Delta \bar{\rho}$. 
As mentioned before, the value of $\Delta$ is typically taken to be 200, as in PTHalos, but  
other numbers are also motivated by the spherical collapse model and N-body simulations.

Once we have the typical velocity dispersion of a halo we assign positions and velocities 
to its galaxies in the following way. 
If there is only one galaxy, we place it at the centre of mass with the velocity of the halo. 
If there is more than one galaxy, the first one is placed at the centre of mass, and 
the others following the NFW density profile. For these galaxies their velocities have
two components: the velocity of the halo centre of mass, 
and a contribution from the velocity dispersion. 
We take the latter to be drawn from a Gaussian distribution with zero mean and variance equal to 
\begin{equation}
\langle v_{1D}^2 \rangle= \frac{1}{3} \langle v_x^2+v_y^2+v_z^2 \rangle  = \frac{1}{3} \langle v^2\rangle. 
\label{eq:vel}
\end{equation} 


\subsection{Redshift Space Distortions}
\label{ssec:rsd}


We use the velocity of galaxies to simulate the effects of Redshift Space Distortions (RSD). 
We therefore alter the positions of galaxies such that each galaxy is set to where it would be observed
in redshift-space coordinates. To achieve this one must convert velocities into displacements
by dividing the former by $\mathcal{H}=\dot{a}=Ha$ and projecting the result in the line-of-sight.
The displacement $\Delta s$ in Mpc/h that corresponds to a velocity of magnitude of 
$\sqrt{\langle v_{1D}^2 \rangle}$ is easily computed. Since $G=3 H_0^2 / 8\pi \rho_{crit}$, 
$\rho_{crit}=\Omega_M^0 \bar{\rho}$, and defining the Hubble expansion rate as $H(z)\equiv H_0 E(z)$, 
one gets

\begin{equation}
\Delta s = \frac{R_{vir}}{E(z)a} \sqrt{F(c)} \; , 
\end{equation}

\noindent where $F(c)$ has been defined in Equation \ref{eqFc}. 
In our PTHalos galaxy mocks we do not use the distant observer approximation (usually 
implemented in the simulations by displacing the particles only in one axis). Instead,
we add the RSD along the line of sight, as it happens in observations.

\subsubsection{Extending the model}

We made several simplifying assumptions to implement the 
the galaxy population method presented in this paper. 
Many effects of the complex relation between halos, matter 
and galaxies are not included in these mock galaxy catalogs. 

We chose to model the galaxies on top of a static realisation of
the matter field, which assumes the evolution over the redshift 
range is small.  This will impact the clustering of matter, as well 
as the associated halo masses. Although we expect this effect to 
be small for the mock galaxy catalogs used in CMASS DR9 results, 
we could improve on the method for future applications and model
this evolution. 

For simplicity, the mocks also neglect any evolution to populating 
dark matter halos, or varying the galaxy bias with redshift.  While 
the sampling of galaxies is adjusted to match the density as a function 
of redshift (see Fig.~\ref{nzfig}), a change in number density is likely 
to correspond to a variation in galaxy selection, and therefore, the 
associated galaxy bias (more luminous galaxies typically correspond to 
lower number densities and higher bias values). Again, we expect a small 
impact on any CMASS DR9 results (Anderson et al. 2012) since much of 
the modeling assumes an average bias value over the redshift range, which 
the galaxy mocks appropriately match. 

We also did not include assembly bias effects 
(Sheth \& Tormen 2004, Croft et al. 2011) in our mocks,
but kept the concentration parameter and HOD independent 
of the environment. For simplicity we also have  set
independent scatters for the number of galaxies in a halo 
and the concentration parameter, even if at a fixed halo
mass they might be related. 

Halos in the mocks are spherical. In reality, as shown by
N-body simulations, they have a range of shapes  
that are correlated to the morphology of the surrounding environment 
(Smith \& Watts 2004, Schneider et al. 2011, White et al. 2010). 
The mocks described in this paper included none of these effects. 
In future versions, a correlation with the environment
could be introduced via the 2LPT estimation of the tidal field. 

Finally, the galaxies in the mocks have no individual colors 
or luminosities. One could include them by following a similar
prescription to one described in Skibba et al. (2006, 2009)
which was constrained by SDSS luminosity and colour dependent
clustering, number densities and colour-magnitude distributions.

\section{Galaxy mocks for the CMASS DR9 sample}

\subsection{Fit to CMASS galaxies}
\label{sec:fit}

\begin{figure}
\center
\includegraphics[width=70mm]{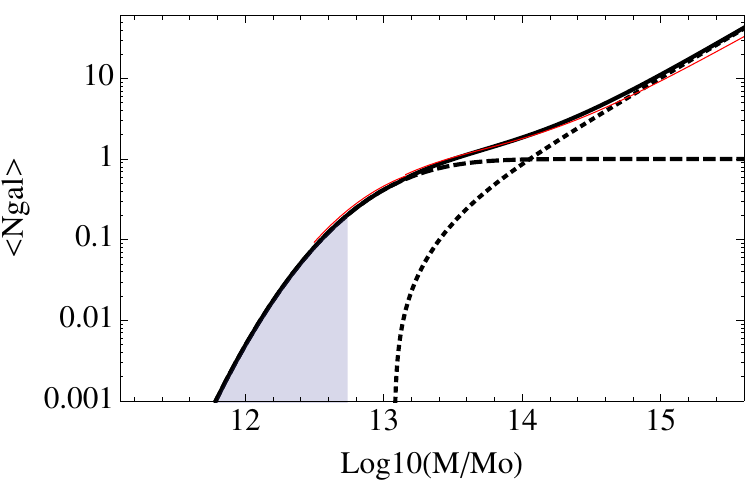}
\caption{Best fit HOD of the mocks (black solid line), with its contribution
split between central galaxies (dashed line) and satellite galaxies (dotted line). 
Grey shadowed area shows the mass range for which galaxies are drawn from matter
particles. White et al. (2011) best HOD fit to CMASS data is shown in red.}
\label{HODplot}
\end{figure}

In order to find values of HOD parameters for the mocks of BOSS CMASS DR9 galaxies, 
we minimize $\chi^2$ between measured BOSS clustering of the full CMASS DR9 sample (NGC plus SGC) 
and the clustering of a mock realization. We choose the mock realization for which the 
power spectrum is closest to the mean of the mocks and compute, for each HOD iteration,
$\xi(s)$ with $s$ between $30$ and $80$ Mpc/h, in an area of a quarter of the sky, 
with a simple mask and a constant $n(z)$, but including redshift space distortions. 
We populate halos until a minimum mass threshold of $M=0.47\cdot10^{13}M_{\odot}/h$, 
which corresponds to halos of 10 particles. The 
rest of the galaxies, which according to each HOD would belong to halos with a lower mass, 
are placed on randomly selected dark matter particles. 

To find values of HOD parameters that minimize  $\chi^2$ we use the simplex
algorithm of Nedler and Mead (1969). We start by making an initial
guess about the values of the HOD parameters and then construct a five-dimensional
simplex with vertices at this initial point and five other points that resulted from
stepping along each coordinate axes with a certain step-size. The algorithm
finds the vertex with the worst $\chi^2$ value and moves it by a combination of
reflection, reflection followed by expansion and multiple contractions until the
value of $\chi^2$ at that vertex is no longer the worst. The algorithm then
keeps contracting the simplex by moving the next worst vertex until the size of the average distance
from the centre of the simplex to its vertices is smaller then a desired
level of accuracy. If the $\chi^2$ surface is unimodal this algorithm is
guaranteed to find the minimum with any desired accuracy.

Our initial guess of HOD parameters was the best-fit set computed using the clustering and
number density of an earlier CMASS sample (see White et al. 2011). After about $40$ steps the resulting best-fit HOD 
was  
\begin{eqnarray}
&&{\rm{log}}(M_{min})=13.09\nonumber \\ 
&&{\rm{log}}(M_1)=14.00\nonumber \\ 
&&{\rm{log}}(M_0)=13.077\nonumber \\
&&\sigma_{{\rm log}M}=0.596 \nonumber \\
&&\alpha=1.0127. 
\end{eqnarray}
We find $\chi^2=5.89$ with 9 degrees of freedom. 
In Figure \ref{HODplot} we show in black the mean number of galaxies 
as a function of halo mass for our best fit. In red we show the best-fit model of 
White et al. (2011). Both agree to within the 1-sigma errors, and the mean number
of galaxies at a given mass, $N(M)$, agrees better than 10 per cent for halos
below $10^{14.5} M_\odot/h$ and better than 20 per cent between $10^{14.5}$
and $10^{15} M_\odot/h$.

The shadowed area in the plot denotes the masses for which we have no halos in the simulation.
The galaxies corresponding to those halos have been assigned 
positions and velocities of randomly selected dark matter particles. 
They form $\sim 25$ per cent of the total of mock galaxies. 
If we did not include them then we would not have recovered a sensible HOD
because we would have had to populate the available low mass PThalos with far too many
galaxies in order to reduce the bias. 

It is possible to set the HOD parameters of the mocks more accurately by  
fitting both the two and the three point correlation functions, as the
latter helps to break degeneracies between the parameters \citep{Sefusatti05,Kulkarni07}.
However, computing the three point function in each step of the fitting 
process is computationally very time consuming. We leave this improvement 
as a possibility for future versions of the mocks. 

\subsection{Geometry and mask}
\label{sec:geometry}

We wish to create mocks with a geometry appropriate for the BOSS CMASS
DR9 galaxy sample, including both the Northern and Southern Galactic caps, with
redshifts between 0.43 and 0.7.  These are the data used in a number
of recent cosmological analyses (\citealt{boss12}, Sanchez et
al. 2012; Samushia et al. 2012; Reid et al. 2012; Nuza et al 2012; Tojeiro
et al. 2012a,b; Ross et al. 2012). In this section we show how we match the DR9 CMASS
geometry.

\begin{figure}
\center
\includegraphics[width=33mm]{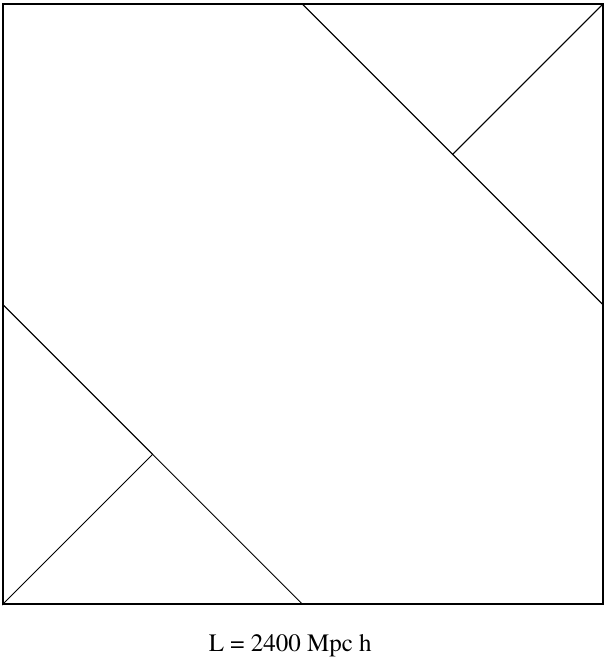}\\
\includegraphics[width=50mm]{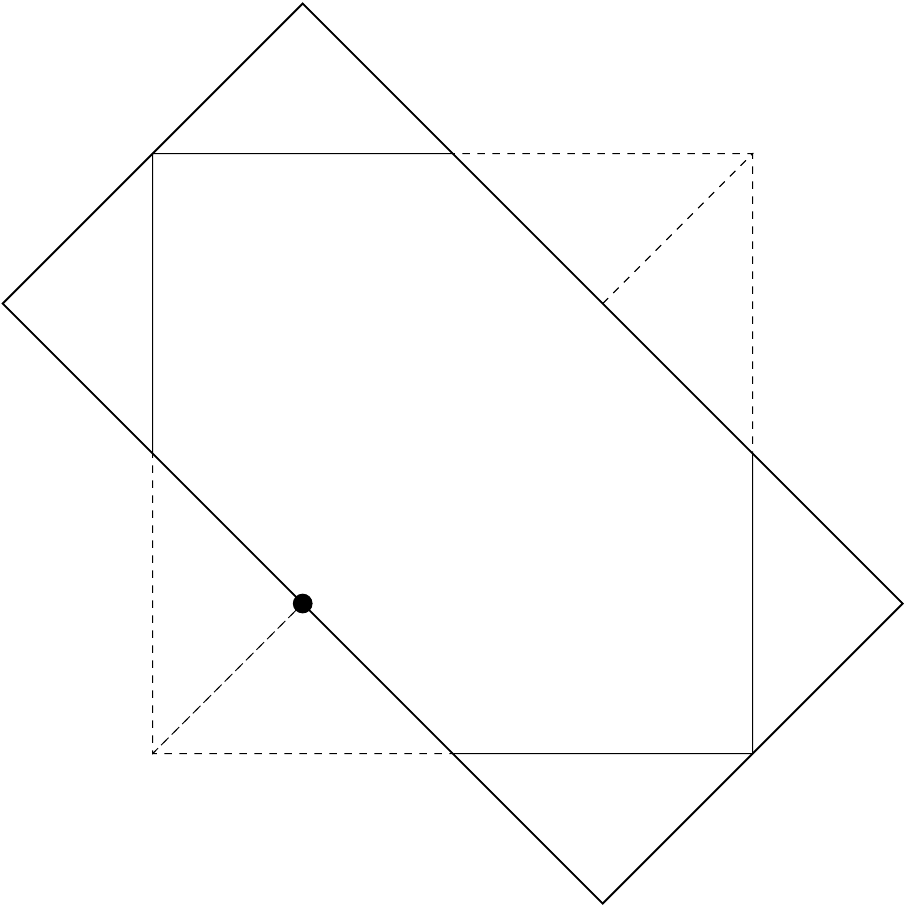}
\includegraphics[width=50mm]{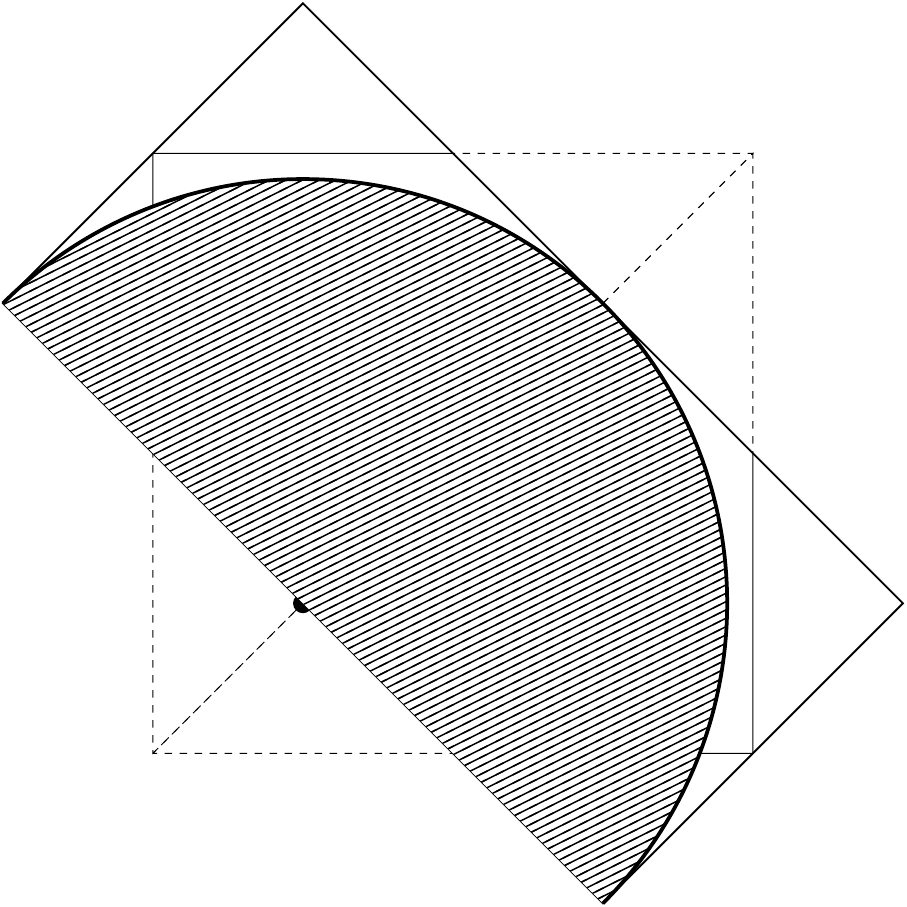}
\includegraphics[width=53mm]{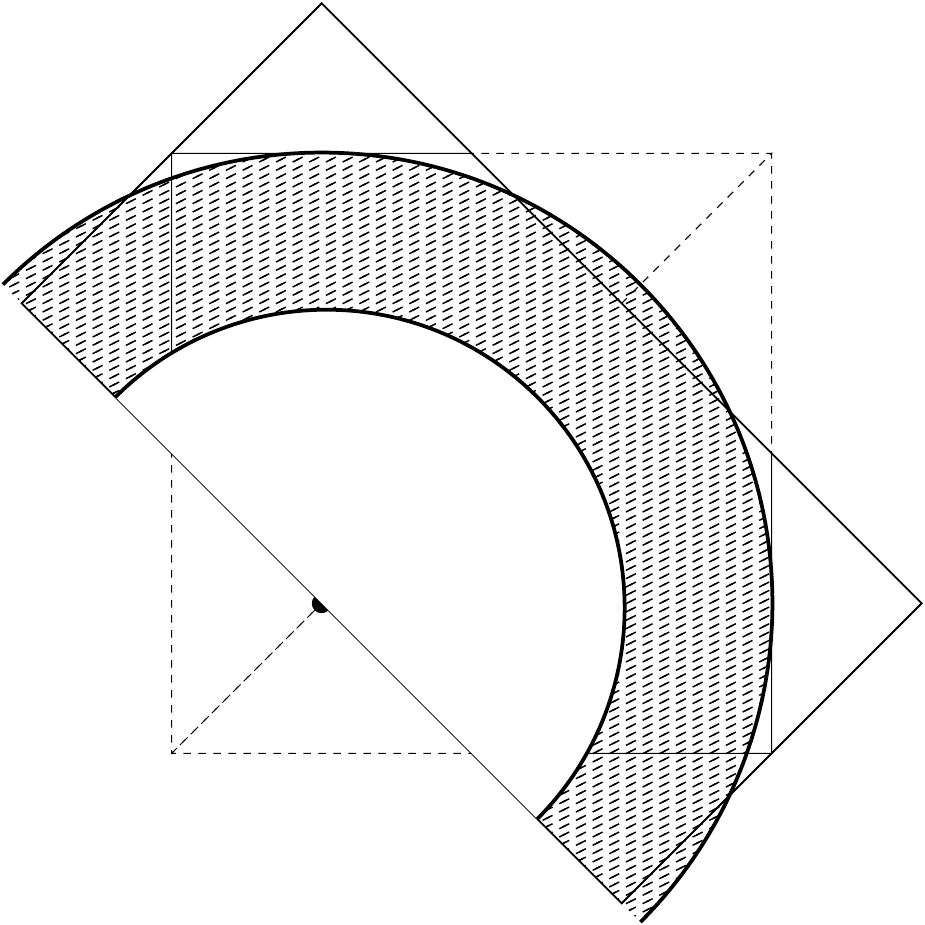}
\caption{
Procedure to fit the geometry of DR9 into the simulation box using periodic boundary conditions. See text for details.} 
\label{bossDR9area}
\end{figure}

The Northern and Southern Galactic cap regions can either be fitted into a
reshaped box with size ${\rm{L}}=2.4{\rm{Gpc/h}}$, which is the size we
adopt for our PTHalos runs. The reshaping is achieved as
follows. Starting with a cubical box of size L, we cut the $xy$-plane
as indicated in the top panel of Figure~\ref{bossDR9area}. Using the
periodicity of the PTHalos simulation we can copy or move the
particles from outside the range ${\rm{L}}/2 < x+y < 3{\rm{L}}/2$ into
that same range.  Thus, as shown in the second panel from the top of Figure
\ref{bossDR9area}, we can obtain a rectangular box of size
${\rm{L}}/\sqrt{2},2{\rm{L}}/\sqrt{2},{\rm{L}}$. The last dimension is defined as the $z$-direction. This technique is similar to
volume remapping of Carlson and White (2010).

With this geometry, placing our observer at
$(x,y,z)=({\rm{L}}/4,{\rm{L}}/4,0)$ we can cover a quarter of the sky
up to a distance of ${\rm{L}}/\sqrt{2}$ from the observer without
repetition of the underlying matter distribution.  This is shown in the third panel from the top 
of Figure \ref{bossDR9area}. For the WMAP cosmological model
this distance is equivalent to reaching a redshift $z=0.663$. Notice,
however, that the constraint of a maximum distance of
${\rm{L}}/\sqrt{2}$ is set only because of the geometry of the $z=0$
plane. Keeping the observer in the same place, but looking into a
direction off the plane, we could go to a higher distance without
hitting repeated volumes. Translating to consider an angular region,
the above maximum distance is only valid if we require a full
180-degree wide view and, for example, for an opening of 126.87
degrees centered on the direction $\hat{e}=(\hat{x}+\hat{y})/\sqrt{2}$
would allow us to reach a distance of $\sqrt{5/8}{\rm{L}}$ without
repetition. It is clear that the actual maximum distance achievable
with any given box without repetition will depend on the angular mask
of the survey being analysed.

To generate the mocks for DR9 CMASS, we first produce a redshift shell
such as that shown in the bottom panel of Figure~\ref{bossDR9area}. We
then rotate the 3D coordinates to fit either the NGC or SGC
angular footprint into the box containing the redshift shell.  Images
of these angular footprints are shown in Figure~\ref{bossmask}. The
extent of these masks means that our boxes are of sufficient size that
mock catalogues containing galaxies with redshifts $z<0.7$ do not
suffer from any repetition of the underlying density field.

In order to mimic the observations as closely as possible, we 
use the {\sc MANGLE} software (Swanson et al. 2008) to differentiate
between sectors that have different observational properties, as
described in Ross et al. (2012). The completeness in the mock galaxies
is defined slightly differently from that of the CMASS DR9 catalogues.
As we are only interested in
large-scales, we do not mimic the full small-scale BOSS targeting
procedure in the mocks. In particular, we ignore the effect of
missing close-pairs of galaxies that result from the fact that we cannot observe two targets closer than 62'' with
the same plate; this is a physical limitation imposed by the size of the fibres. We also ignore the effect of plate-scale angular
variations in our redshift success rate. In Section~3 of
\citet{boss12} two completeness measures are defined: the fraction of objects
targeted that are observed or are in a close-pair, $C_{\rm BOSS}$, and the
fraction of galaxies with good redshifts, $C_{\rm red}$. For the
mocks, we revise the definition of sector completeness such the
angular variations in galaxy density follow those in the sample with
good redshifts, ignoring close-pairs. We therefore define
\begin{equation}  \label{eq:Cmock}
  C_{\rm mock}=\frac{N_{\rm obs}}{N_{\rm targ}-N_{\rm known}},
\end{equation}
where $N_{\rm obs}$ is the number of objects observed
spectroscopically by BOSS in any sector, $N_{\rm targ}$ is the number
of target objects, and $N_{\rm known}$ is the number that already have
good-quality known redshifts. Following \citet{boss12}, the redshift
completeness is defined as
\begin{equation}  \label{eq:Cred}
  C_{\rm red}=\frac{N_{\rm gal}}{N_{\rm obs}-N_{\rm star}},
\end{equation}
where $N_{\rm gal}$ is the number of targets within a sector, observed
by BOSS and subsequently spectroscopically classified as galaxies with
good redshifts, and $N_{\rm star}$ is the number classified as stars.
We subsample galaxies in our mock catalogues based on the product
$C_{\rm mock}\times C_{\rm red}$. i.e. we subsample based on angular
fluctuations of galaxies with good redshifts, ignoring other
subtleties. The implemented angular mask can be seen in Figure
\ref{bossmask}.

As we are only interested in matching the large-scale clustering
signal we do not include small-scale holes in the
survey mask such as those due to SDSS fields with know photometric
problems, objects observed with higher priorities, bright stars, and
plate centres (see \citealt{boss12} for details). In total these
remove 5 percent of the mask area, as defined by overlapping tiles,
and the holes represent small angular patches that are approximately
randomly distributed. As we are only interested in large-scales, the
net effect on removing such holes is equivalent to reducing the galaxy
density, rather than the volume. Consequently, we simply match the
total galaxy number after removing these regions from the CMASS DR9
galaxy catalogue.

In order to mimic the measured redshift distribution we
subsample the galaxies in each PTHalos mock based on a smooth fit to
the measured redshift distribution, $n(z)$. We do this separately for
the NGC and SGC areas, as they have slightly different
redshift distributions (see Fig.~\ref{nzfig}, and Ross et al. 2012).

Using the above procedure we generated 600 PTHalos mocks with
WMAP underlying cosmology. 

\section{Results from the CMASS DR9 mocks}
\subsection{Correlation Function Monopole}
\label{sec:monopole}

\begin{figure}
\center
\includegraphics[width=80mm]{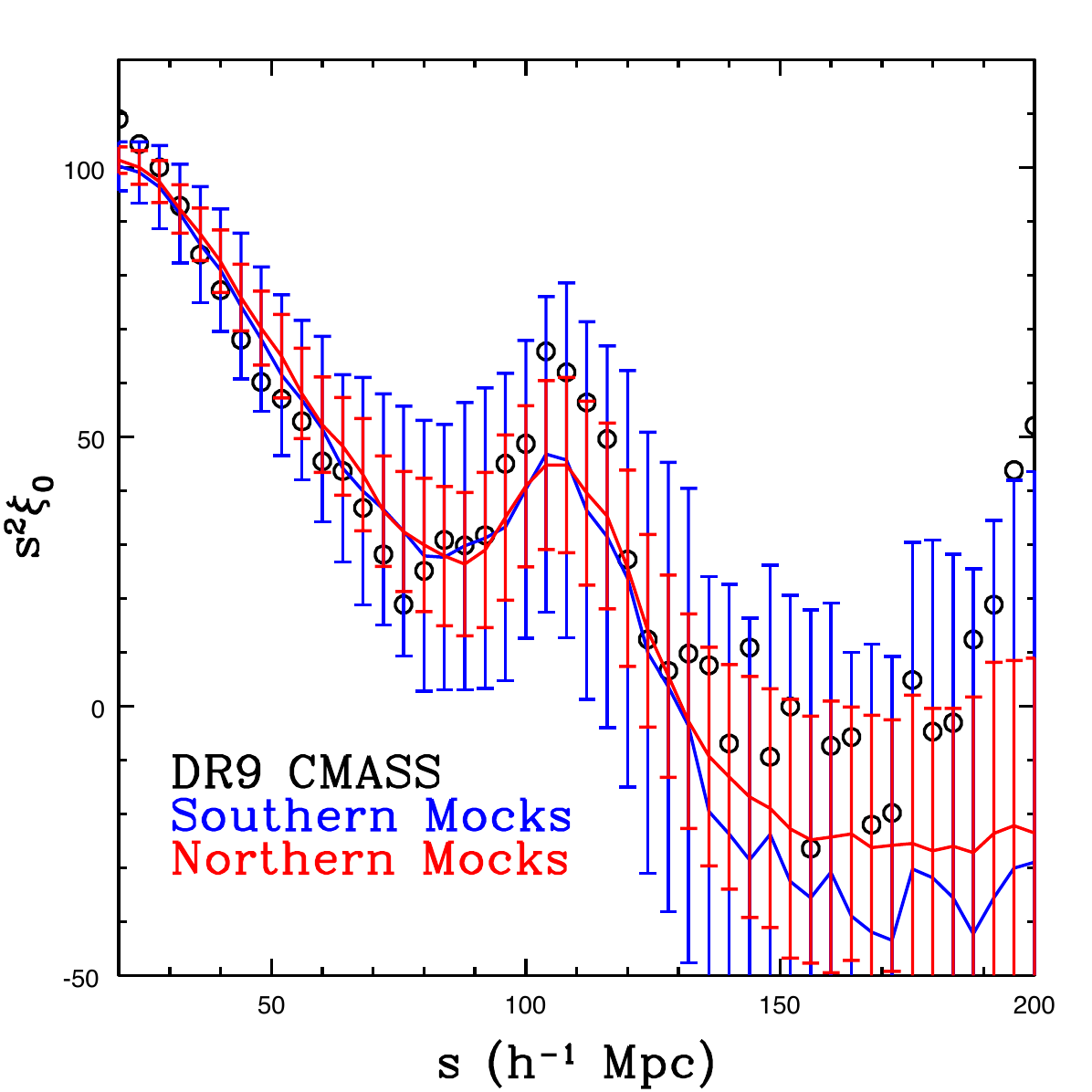}
\includegraphics[width=80mm]{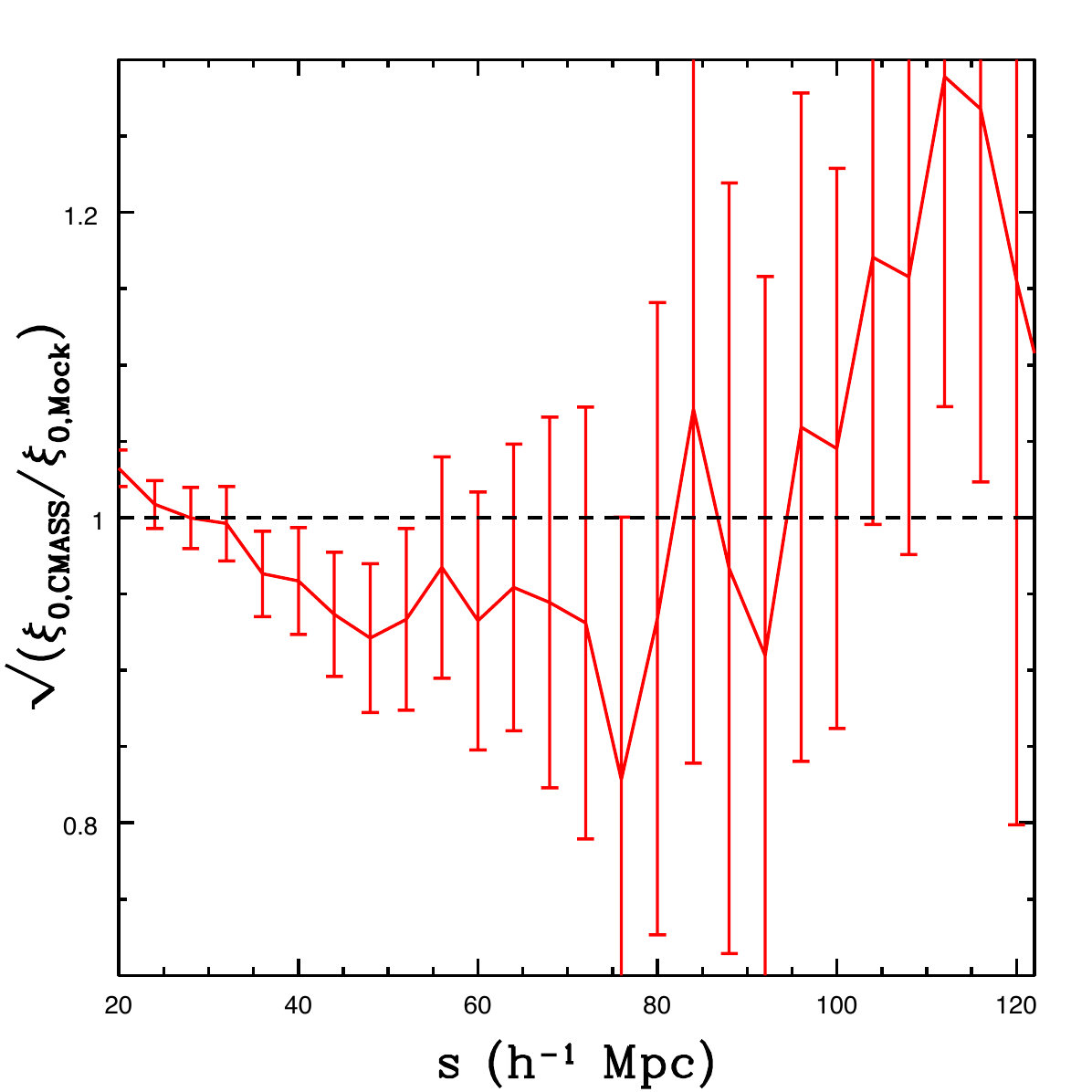}
\caption{ TOP: Correlation function monopole $\xi(s)$ of the NGC and SGC mocks, respectively 
shown in red and blue. The NGC footprint, having larger area has smaller errors. CMASS DR9 data is shown in open circles. Error bars are from the 600 galaxy mock catalogues. BOTTOM:
The relative bias between the mocks and the data, shown for the NGC mocks.}
\label{xiDR9}
\end{figure}

We used the Landy and Szalay (1993) estimator to calculate the anisotropic redshift space 
correlation function, $\xi(s,\mu)$, where $s$ is the redshift-space separation and $\mu$ 
is the cosine of the angle between the galaxy pair and the line-of-sight:

\begin{equation}
\xi(s,\mu) = \frac{DD(s,\mu)-2DR(s,\mu)}{RR(s,\mu)}+1 \; ,
\end{equation}

\noindent where $D$ stands for the data number counts 
and $R$ stands for the random sample number counts with the same redshift distribution 
and angular footprint as the data sample. 

The moments of $\xi(s,\mu)$, expanded in Legendre polynomials, contain all of the information
about the correlation function. They are given by 
\begin{equation}
\xi_{\ell}(s) =  \frac{(2\ell + 1)}{2}\int_{-1}^{1} \xi(s,\mu) P_{\ell}(\mu) {\rm d}\mu \; .
\end{equation}

We will focus on the  monopole $\xi_0$ and the quadrupole $\xi_2$ (see below) 
as in linear theory they contain most of the information.  
We weight pair counts based on their number density,
with weights $w=( 1+n(z) P_{fkp})^{-1}$ (Feldman, Kaiser \& Peacock 1994) where
$P_{fkp}=20000. h^{-3} Mpc^3$ The same applies to the power spectrum. For more details on the
weighting see Ross et al. (2012) and \citet{boss12}.     

In the top panel of Figure \ref{xiDR9} we present the mean of the monopole of the 
correlation function $\xi_0(s)$ from our mocks. 
The red and blue lines show the mean of the 600 mocks using the NGC and SGC footprint
respectively. The two means are similar as expected, and differ only because of cosmic
variance and differences in the survey geometry.  
The error bars show the RMS of the mocks,
and thus give an estimation of the typical dispersion between them. The errors
are smaller for the NGC because of the larger area. The DR9 CMASS $\xi_0(s)$ is shown as open circles.  

The relative bias between the data and the mean of the NGC mocks 
is shown in the bottom panel of Figure \ref{xiDR9} 
The differences between data and mocks are within 10 per cent for most of the scales below 
100 Mpc/h.  

\begin{figure*}
\begin{minipage}{7in}
\center
\includegraphics[width=180mm]{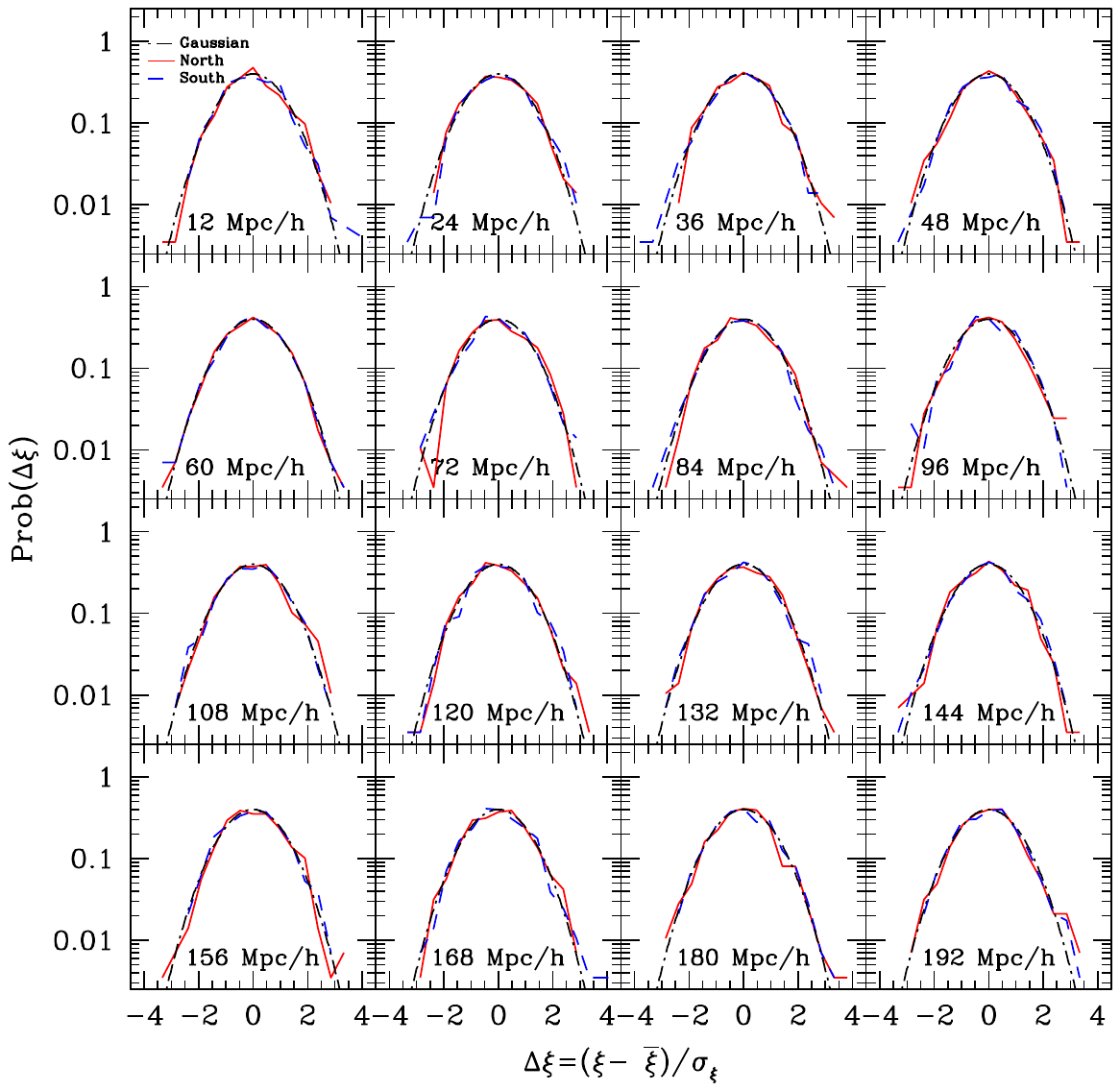}
\caption{ Histogram of the normalised residual counts of the correlation 
function $\xi(s)$ split into scales from $12$ to $192$ Mpc/h.} 
\label{xidist1d}
\end{minipage}
\end{figure*}

\begin{figure*}
\begin{minipage}{7in}
\center
\includegraphics[width=180mm]{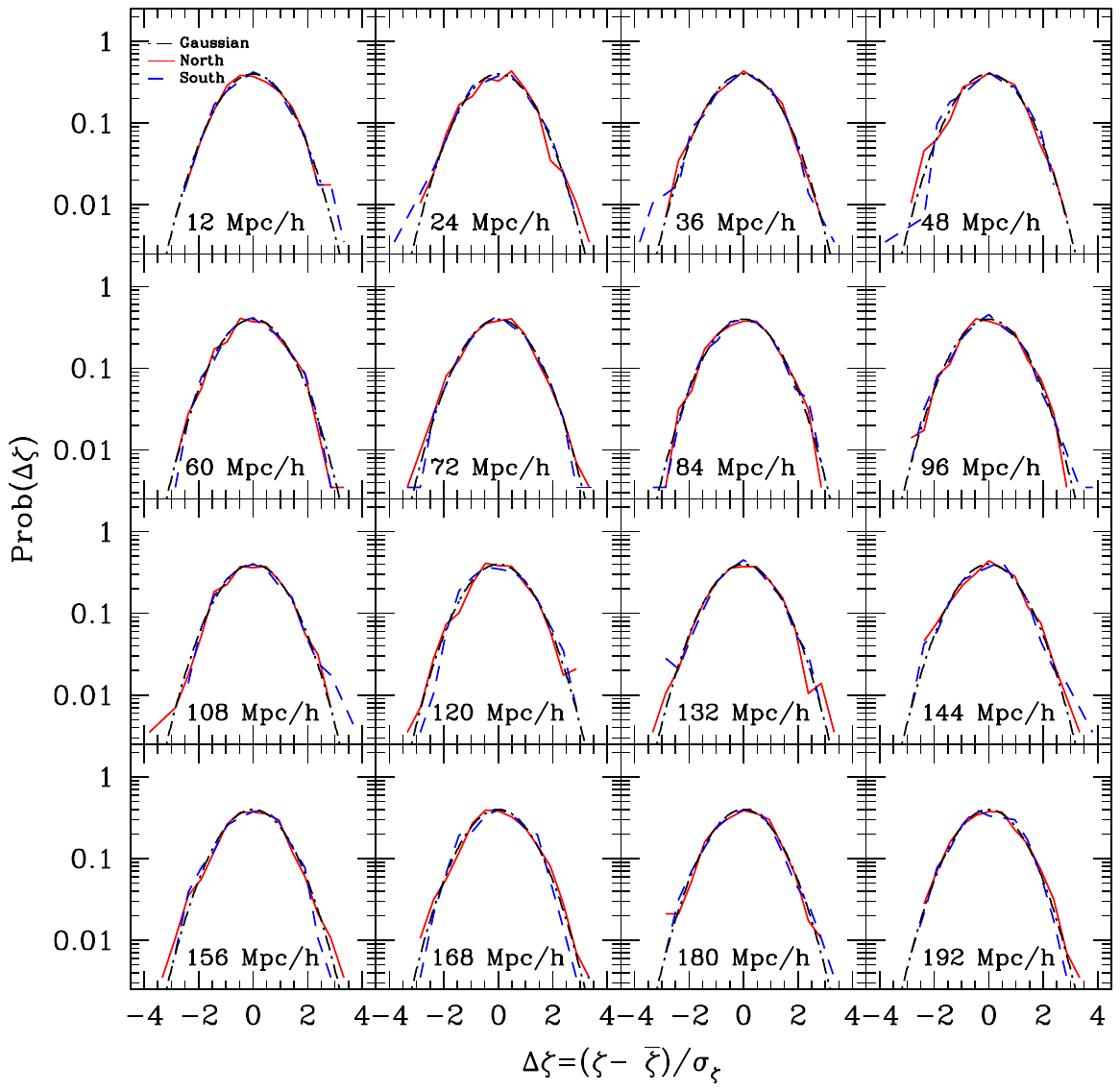}
\caption{ Histogram of the normalised residual counts of the correlation 
function $\xi(s)$ after being projected into the space where the covariance is diagonal. 
Results split into scales from $12$ to $192$ Mpc/h.} 
\label{xidistdiag1d}
\end{minipage}
\end{figure*}

In Figure \ref{xidist1d} we present the distributions of the values of the correlation function 
of the mocks for several separation distances, in normalised units. 
That is, for each bin in $s$ of the correlation 
function $\xi(s)$ one can compute its variance and express the value of the 
correlation function in its units. The histogram of the 600 values is also
normalised to one. Thus if the mocks are Gaussian this distribution should 
follow a normalised Gaussian distribution. In red solid lines we show the results for
the NGC sample, and in blue dashed lines the results for the SGC sample. We see no
significant deviation from the Gaussian distribution shown in black dotted lines, and there
is no particular scale appearing to perform worse than the others.  

The values of the correlation functions at different scales are correlated. 
To have a better understanding of their distribution we have made a 
transformation of the correlation function into the basis where the
covariance matrix is diagonal. In Figure \ref{xidistdiag1d} we show
the normalised distributions of the transformed correlation functions $\xi_t(s)$,
at different scales $s$. In this basis the distribution in each scale 
is independent of the others. In red solid lines we show the results for
the NGC sample, and in blue dashed lines the results of the SGC sample. We see no
significant deviation from the Gaussian distribution shown in black dotted lines, and,
again, there is no particular scale appearing to perform worse than the others.

To check the compatibility of the distribution of the mocks  
with a Gaussian distribution we performed a Kolmogorov-Smirnov test on the
measured distribution function of $\xi_t(s)$ of the NGC sample. 
The result depends on the range of scales used. For scales between 
$50 < s < 150 $ Mpc/h in 9 per cent of the cases, a sample
drawn from a Gaussian distribution with zero mean and unit variance would 
appear less Gaussian than that the distribution obtained from the 600 mocks. 

\subsection{Correlation Function Quadrupole}
\label{sec:quadrupole}

\begin{figure}
\center
\includegraphics[width=80mm]{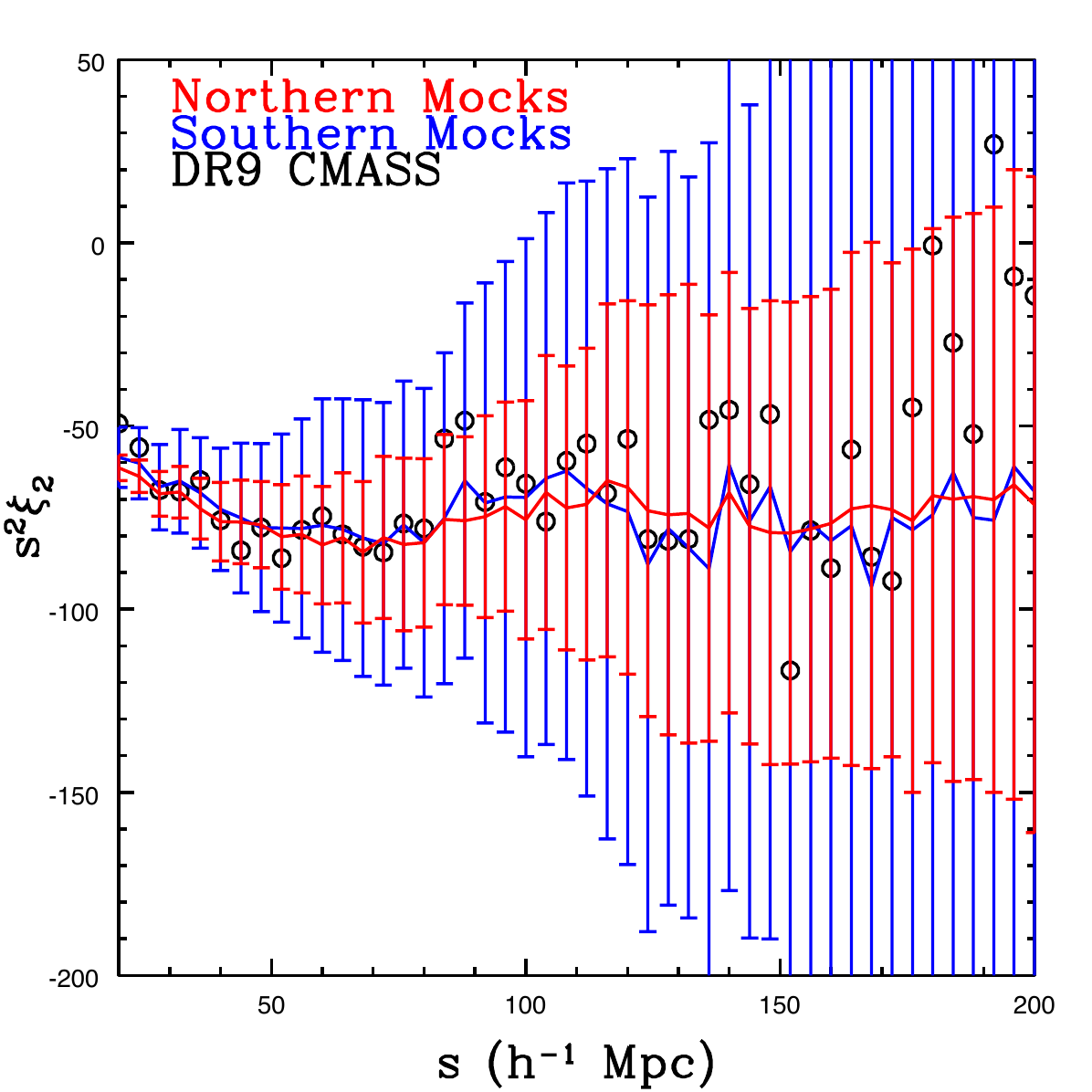}
\includegraphics[width=80mm]{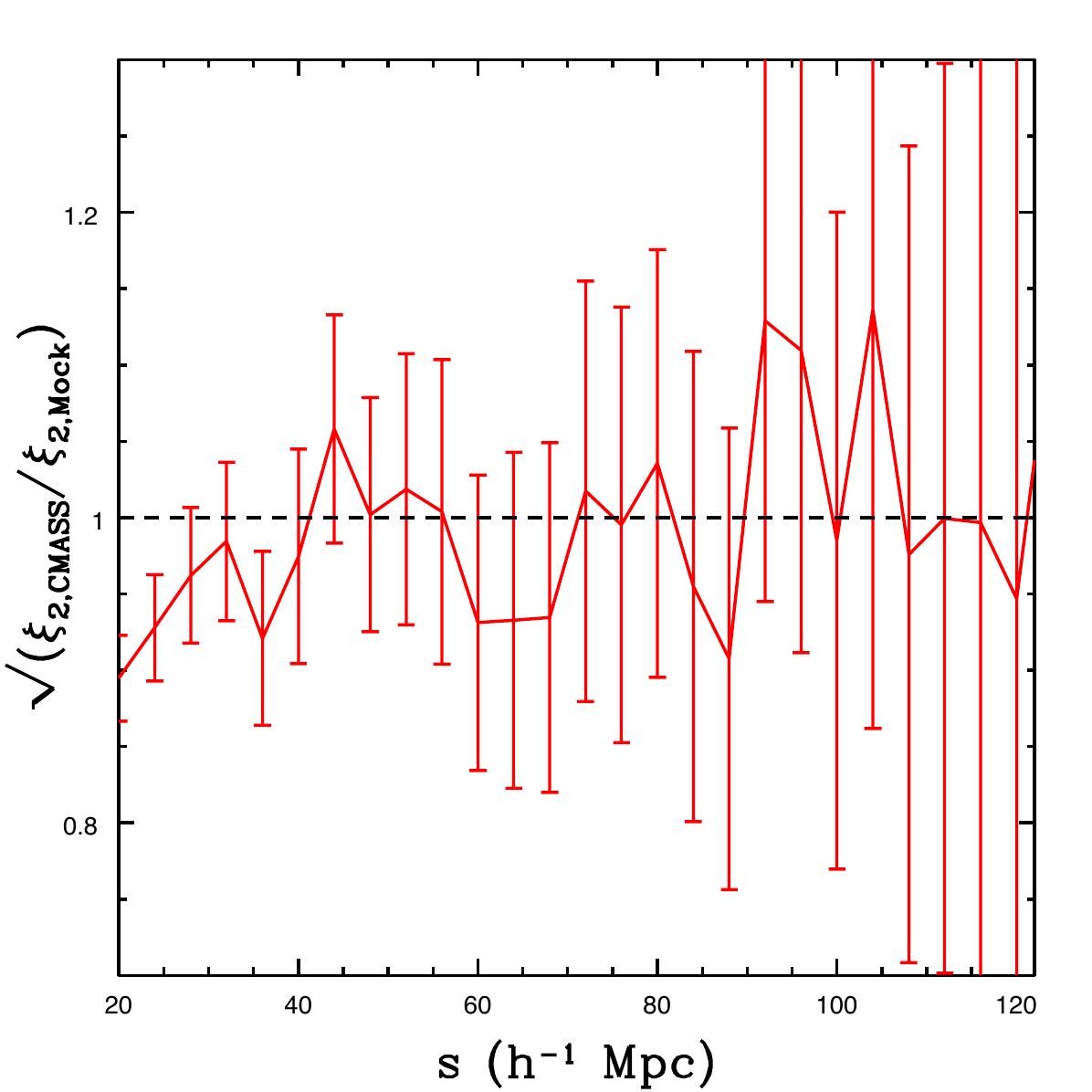}
\caption{ TOP: Correlation function quadrupole $\xi_2(s)$ of the NGC and SGC mocks, respectively 
shown in red and blue. The NGC footprint, having larger area has smaller errors. CMASS DR9 data is shown in open circles. Error bars
show the RMS of 600 galaxy mock catalogues. BOTTOM:
The relative bias between the mocks and the data, shown for the NGC mocks.}
\label{fig:xi2}
\end{figure}

In Figure \ref{fig:xi2} we show the average measurement of the quadrupole for the NGC (red) and SGC (blue) mocks.
The quadrupole measured from the CMASS DR9 data is shown in the open circles. 
Error bars show the RMS of the 600 mocks.  
The anisotropic clustering, i.e., the quadrupole, can be used to estimate 
the growth rate of structure $f$. 

In the linear regime the expression for the redshift space distortions is \citep{Ham92}
\begin{align}
  \xi_0(s)& = & \left(b_g^2+\frac{2}{3}b_gf+\frac{1}{5}f^2\right)\xi^{\rm r}(s)\; ,\\
  \xi_2(s)& = & -\left(\frac{4}{3} b_g f +\frac{4}{7} f^2 \right)\left[\bar{\xi}(s) - \xi^{\rm r}(s)\right]\; ,
  \label{eq:xi02}
\end{align}
\noindent
where $\xi^{\rm r}$ is the real space matter correlation function normalised so that 
\begin{equation}
  \displaystyle\int_0^\infty\xi^{\rm r}(s)s^2ds=1 \; , 
  \label{eq:xirnorm}
\end{equation}
\noindent
$\bar{\xi}$ is given by
\begin{equation}
  \bar{\xi}(s) = \frac{3}{s^3}\displaystyle\int_0^s \xi^{\rm r}(s')s'^2ds'\; ,
  \label{eq:xibar}
\end{equation}
\noindent
and $b_g$ is the bias of galaxies.

We have estimated values of galaxy bias $b_g$ and growth rate $f$ in the mocks
by performing a joint fit to the measured redshift space monopole and quadrupole 
of the correlation function between the scales of $50 < s < 150 {\rm Mpc/h}$. 
We used the standard perturbation theory predictions of the real space pairwise 
halo velocity statistics to model the non-linear contribution to the redshift
space correlation function \citep{Reid2011}. The fit results in $b_g = 1.90$ and $f = 0.729$.
The value of the 
growth rate recovered in this fit is very close to the value from general relativity
for this cosmological parameters, $f=0.744$ (only a 2 per cent difference). 

Notice that if we were to fit the quadrupole of the correlation function using
only the linear theory to model the shape of the multipoles and the linear Kaiser
formula for redshift-space distortions (Eq \ref{eq:xi02}), then the recovered best
value of the fit to $f$ would be lower. This is expected due to non-linearities, 
which act to decrease the redshift-space anisotropies predicted by the Kaiser formula, 
even on relatively large scales \citep{Scoccimarro04}.

\subsection{Power spectrum}
\label{sec:power}

\begin{figure}
\center
\includegraphics[width=80mm]{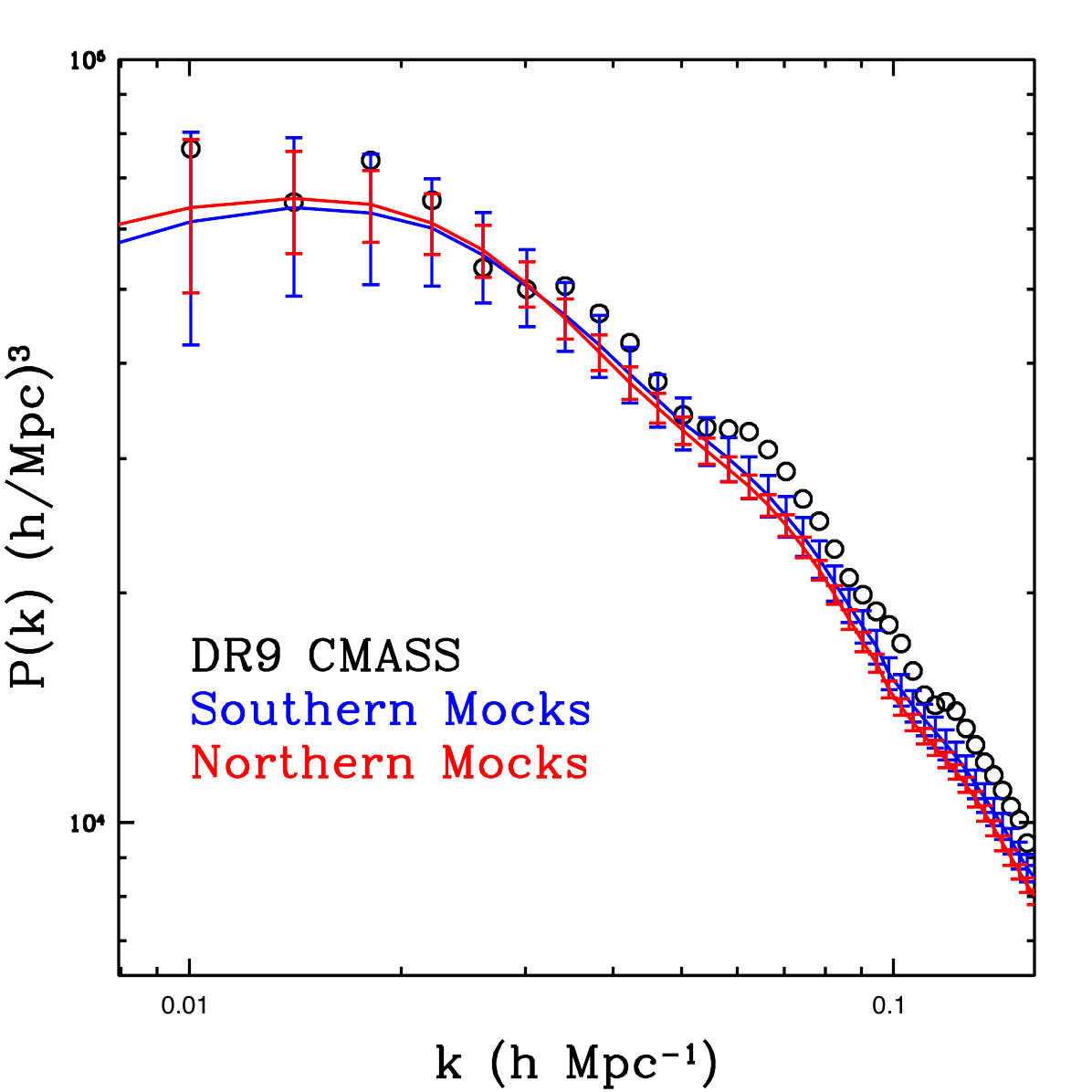}
\includegraphics[width=80mm]{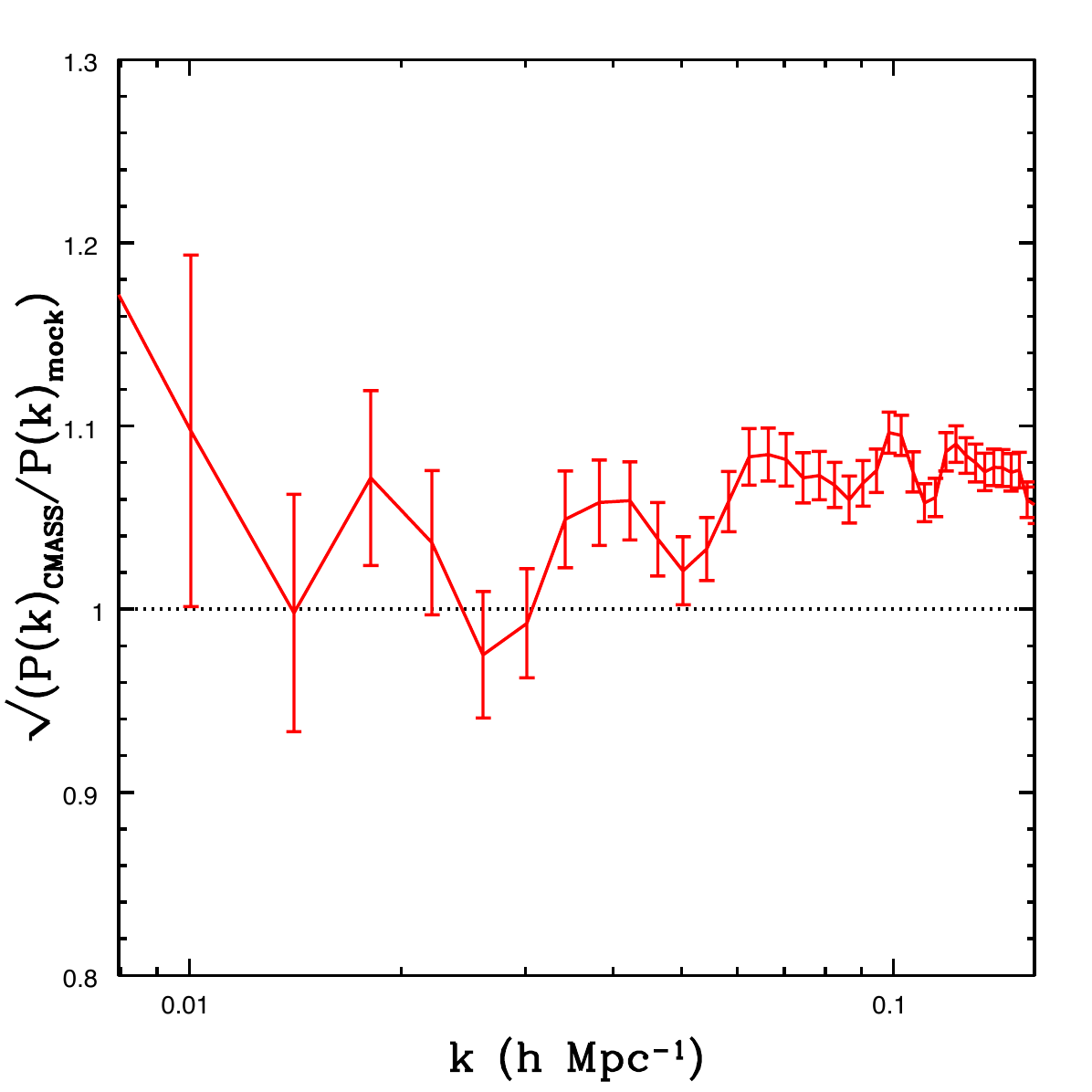}
\caption{ TOP: Power spectrum $P(k)$ of the NGC and SGC mocks, respectively 
shown in red and blue. CMASS DR9 data is shown in open circles. Error bars are from the 600 galaxy mock catalogues. BOTTOM:
Relative bias between the mocks and the data, shown for the NGC mocks. The NGC footprint has the smaller errors because of its larger area.}

\label{pkDR9}
\end{figure}

The top panel of Figure \ref{pkDR9}show the average power spectrum of the mocks, 
both for the NGC and SGC footprints, compared with the DR9 CMASS galaxy
power spectrum. In the bottom panel we show the relative bias between the data
and the mocks, i.e,  the square root of the ratio between their respective power
spectra. The relative bias is within 10 per cent for scales between $0.01 < k < 0.2$
and increases at very low k. 

The amplitude of the power spectrum of the data is slightly higher
than the average of the mocks. Consequently, the mocks underestimate the errors of the
the amplitude of the measured power spectrum by the same factor. However, if what one wants
to estimate the position of the BAO peak as in \citet{boss12}, then 
and a lower amplitude of the mocks would give conservative larger errors on the peak position. 
Thus, if anything, we are over-estimating our errors. 

We have checked that the power of the mocks at 
these scales have a scale dependence consistent with the theoretical matter 
power spectrum used as an input, convolved with the window function of the survey.

\section{comparison with analytic prediction}
\label{sec:analytical}

In this Section we compare the covariance matrix of the galaxy mocks 
described above to a covariance matrix 
based on an analytical approach of de Putter et al. (2012). This approach
provides a prescription for the dark matter power spectrum
covariance matrix, taking into account the effects of survey geometry
and using standard perturbation theory to include non-linear effects.
The resulting covariance matrix has been shown to agree well with N-body simulations
for modes $k < 0.2 h/$Mpc. However, to analytically describe 
the covariance matrix of the {\it galaxy} two point function,
the effects of galaxy bias, redshift space distortions and shot noise 
need to be taken into account in addition to the dark 
matter prescription. 
We now describe our simplified assumptions for these ingredients below.

Galaxy bias is assumed to be linear and scale independent,
with a value of $b_g=1.9$, which is the best fit to the mocks.
Shot noise due to the finite number of galaxies is incorporated
following \citep{fkp94}, which treats the shot noise as Gaussian.
Finally, redshift space distortions are incorporated using the expression
based on linear theory and the plane-parallel approximation \citep{kai87}
$\delta_g({\bf k}) \rightarrow (1 + \beta (\hat{k} \cdot \hat{n})^2)
\, \delta_g({\bf k})$
where $\beta = f/b_g$, with $f \equiv d\ln d/d\ln
a \approx \Omega_m^{0.55}(z)$ the growth factor,
and $\hat{n}$ the line-of-sight unit vector.
On large scales, this causes a simple rescaling of the covariance matrix
by the angle average of the fourth power of the ``Kaiser factor'',
$a_{\rm rsd}(\beta) \equiv 1 + 4/3 \beta + 6/5 \beta^2 + 4/7 \beta^3 +
1/9 \beta^4$, which we use to multiply the entire covariance matrix.

Putting it all together, the analytic model for the covariance between 
galaxy power spectrum estimators in bins $i$ and $j$ in
$k-$space is obtained by symmetrizing
\begin{eqnarray}
{\bf c}_{ij}^{\rm gal} = \left[ 2 \int_i \frac{d^3 {\bf k}}{v_{k,i}}
\, \int_j \frac{d^3 {\bf k'}}{v_{k,j}} \, |b_g^2 \, p(k) \, q({\bf k}
- {\bf k'}) + s({\bf k} - {\bf k'})|^2 \right. \nonumber \\
\left. +  \, b_g^4 \, {\bf c}_{ij}^{\rm matt, non-lin} \right] \times
a_{\rm rsd}(\beta), \nonumber \\
\label{croland}
\end{eqnarray}
where $v_{k,i}$ is the $k$-volume in a bin i, $p(k)$ is the matter power spectrum, and 
\begin{equation}
q({\bf k}) \equiv {I_{22}({\bf k})\over I_{22}({\bf 0})}, \ \ \ \ \ s({\bf k}) \equiv {I_{12}({\bf k})\over I_{22}({\bf 0})},
\end{equation}
with $I_{ij}({\bf k}) = \int \bar{n}^i({\bf r}) w^j({\bf r}) {\rm e}^{i {\bf k}\cdot {\bf r}} d^3 r$,  $\bar{n}$ is the selection function of the survey and $w=(1+\bar{n} P_0)^{-1}$ the optimal FKP weight function. In Eq.~(\ref{croland}),
${\bf c}_{ij}^{\rm matt, non-lin}$ describes the non-Gaussian matter power spectrum covariance matrix and is given by the second and
third lines of Eq.~(47) in de Putter (2012), to which we refer the reader for more details.

To obtain the covariance matrix of the two-point function, this matrix is
transformed applying the linear transformation between the Feldman Kaiser and Peacock (1994) 
power spectrum estimator and the Landy \& Szalay (1993) correlation function
estimator.

The main caveats in the analytic method come from the simplified transformation described above between the real-space dark matter covariance matrix and the redshift-space galaxy covariance matrix. In reality, the galaxy bias
is not linear and this affects the non-Gaussian contribution to the covariance matrix. Moreover, the analytic model only describes redshift space distortions at the linear level,
and therefore does not include ``fingers of god'' effects which appear already on weakly non-linear scales. Finally, the shot noise also contributes to the non-Gaussian part of the covariance matrix. However, the analytic description is expected to work 
well in the linear regime,
and provides a reasonable estimate to compare to the numerical method from the mocks in the range of scales of interest (35-140
 Mpc/$h$).

\begin{figure}
\center
\includegraphics[width=80mm]{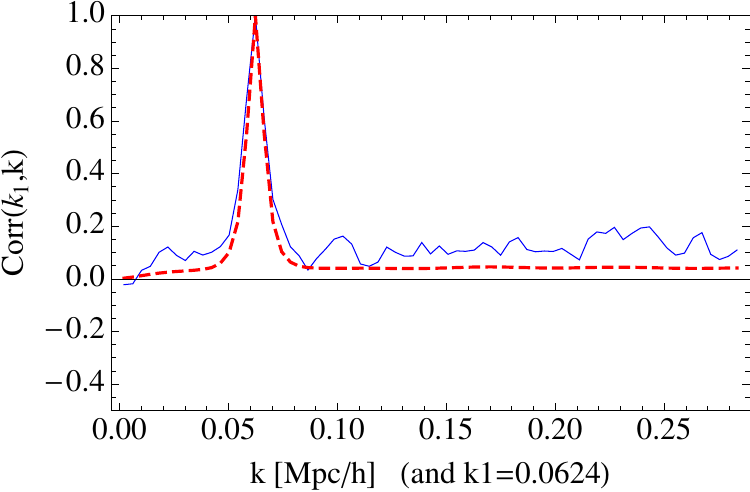}
\includegraphics[width=80mm]{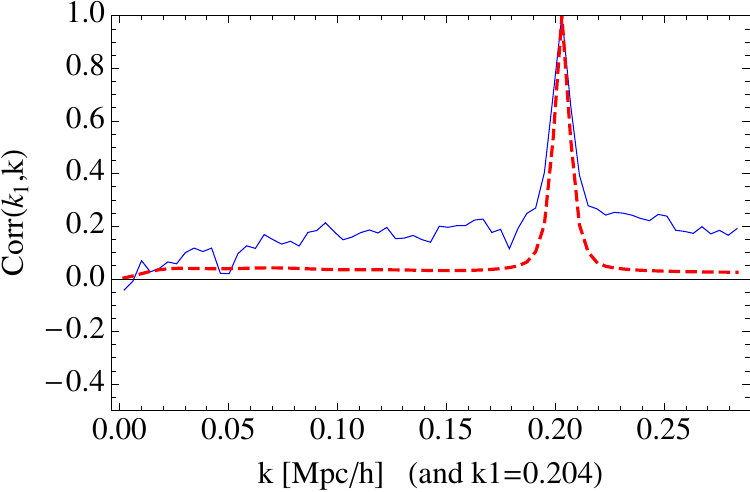}
\caption{ Correlation coefficients $C(k,k_1)$ for the power spectrum of the mocks (in blue solid lines) compared to the analytical values (in dashed red lines)}
\label{fig:corrcoeff}
\end{figure}

\begin{figure}
\center
\includegraphics[width=80mm]{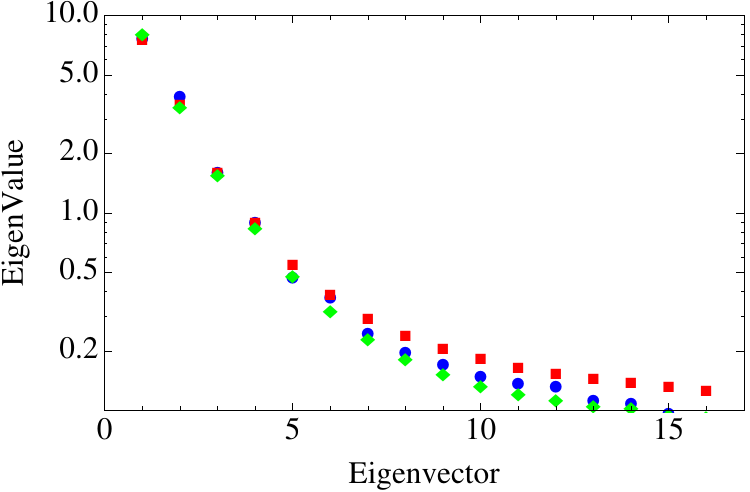}
\caption{ Eigenvalues of the normalised covariance matrix of the mocks' correlation function (blue circles) compared to an smoothed version of it (green diamonds) and to analytical values (red squares).}
\label{fig:eigen}
\end{figure}

We now compare the galaxy mock covariance matrix with the analytic estimates
using the DR9 NGC footprint and assuming our WMAP fiducial cosmology
described in Section \ref{sec:lasdamas}.

We start first with the power spectrum covariance matrix.
Figure \ref{fig:corrcoeff} shows the normalised (to have unit diagonal) covariance matrix or cross-correlation coefficients, $C(k_1,k)$, of the power spectrum
of the mocks (in solid blue lines) and of the analytical model (in red dashed lines). The plots are for the value $k=0.0624~h$/Mpc
and $k=0.204~h$/Mpc but similar results are obtained when fixing $k_1$ at other values. The mocks
have a somewhat stronger correlation amplitude than the analytical model,
which is not surprising given that nonlinear contributions from redshift-space distortions and bias are 
not taken into account in Eq.~(\ref{croland}), as discussed above.

We now turn to configuration space. 
Figure \ref{fig:eigen} shows the eigenvalues from the mock correlation functions (blue circles) compared
with the eigenvalues of the analytical model (red squares). Both give comparable results,
with the four largest eigenvalues having differences at the $\lesssim 10 $ per cent level,
which increases up to 25 per cent for the fourteenth eigenvalue.

Figure \ref{fig:eigen} also shows a comparison with the method of Xu et al. (2012), denoted by green diamonds, which is based on  fitting a modified form of the Gaussian covariance matrix to the sample covariance matrix from the mocks using a maximum likelihood approach. The eigenvalues of the smoothed version of the covariance matrix are consistent at the 10 per cent level with the values of the sample covariance from the mocks.

\section{Correlation function and covariance matrix tables}
\label{sec:covar}

We have created galaxy mocks with the DR9 footprint and CMASS redshift selection.
They can be used by the community as means of computing covariance
matrices for large-scale structure analysis. 

Tables 2 and 3 show respectively the mean of the monopole of correlation function and the covariance
matrix of the 600 mocks for the NGC and SGC footprints. The logarithmic binning of the correlation
function here is the same as in Samushia et al. (2012) and Reid et al. (2012). 
 
Note that the 600 NGC and SGC mocks are obtained
from the same 600 primary PTHalos fields. Therefore the NGC and SGC mocks are not truly independent. 
The measured correlation between the mocks with the same seeds is however small, $(3\pm2)$ per cent. 
Due to slight sample variation between NGC and SGC samples (Ross et al. 2012),
we adopt a different fitted n(z) for both. 

PTHalo mocks, tables of the covariance matrices, and covariance matrices
with different binning will be available from the mocks website
\footnote{http://www.marcmanera.net/mocks/}
after the DR9 is made public and this work is published. Updated version of the mocks
will be also hosted at this site. 


\section{conclusions}

In this paper we have presented a method to quickly produce large number of 
galaxy mocks for large scale structure analysis. 
The method has five steps: 
\begin{enumerate}
\item
Generate a dark matter particle field using 2LPT. 
\item
Obtain halos using a FoF algorithm with an appropriate linking 
length, which we have tested to be b=0.38 times the mean inter-particle 
separation at redshift $z\sim 0.5$. 
\item
Promote the mass of these
2LPT halos to new PTHalos masses using the transformation function
that maps the mean mass 2LPT halo mass function to the desired PThalo mass 
function, typically measured or derived from simulations.
\item
 Populate the halos with galaxies using an HOD prescription with the HOD parameters
fit to reproduce the correlation function of the observed survey, 
in this case CMASS DR9 sample.
\item 
Apply survey mask and galaxy selection criteria.
\end{enumerate}

The time savings compared to doing mock catalogues from N-body simulations comes from the first step (where for the particle numbers used here, 2LPT is about three orders of magnitude faster than N-body simulations). The total time spent in making mock catalogues in PTHalos is dominated by the subsequent steps, and thus the speedup factor at the end of the procedure is reduced to about two orders of magnitude.

We have tested the clustering of the PTHalos generated by this method
by comparing the halo-matter cross-power spectrum of 40 PTHalos realisations
with that of 40 LasDamas N-body simulations with the same cosmology, 
mass resolution and Fourier phases. The clustering is recovered to within 
10 per cent level (see Figure \ref{crosspowermany}). And the correlation 
coefficients show that the PTHalos trace the same structures as the N-body 
simulations (see Figure \ref{covariance40}). 

We have used the LasDamas N-body simulations to test the proper 
linking length value to be used with FoF halos from 2LPT fields.
We have found that the theoretical motivated value of $b\sim 0.38$
(Section \ref{sec:linktheo}) is the one that performs best within
the range of values we test against an N-body simulation
(Section \ref{sec:bcal}). 

We have applied our method to generate 600 galaxy
mocks catalogues for the DR9 BOSS CMASS galaxies. For these
mocks we have fixed the mass function of PTHalos to that of
Tinker et al. (2008), for our cosmology, and set the HOD 
parameters by fitting the DR9 data correlation function 
(see Section \ref{sec:fit}). In Sections \ref{sec:monopole},
\ref{sec:power} and \ref{sec:quadrupole} we present the
the monopole of the correlation function, the monopole
of the power spectrum, and the quadrupole of the correlation function,
and its comparison to the CMASS DR9 data. In Section
\ref{sec:covar} we presented the covariance matrices. 

The 600 mocks were produced using a cubical box reshaped
to match BOSS DR9 geometry, separately for both NGC and SGC footprints.
Redshift space distortions are included. Mocks have been 
used within the BOSS collaboration in the analysis of the  
Baryon Acoustic Oscillations \citep{boss12}, redshift space
distortions (Reid et al. 2012, Samushia et al. 2012), clustering of galaxies
below 100Mpc/h compared with simulations (Nuza et al. 2012), systematics
of CMASS DR9 galaxies (Ross et al. 2012), bias evolution (Tojeiro et al. 2012),
and fit to the full shape of the correlation function (Sanchez et al. 2012). 

Finally, we have compared the covariance matrices to 
analytical covariance matrices and found a good agreement
with differences less than 10 per cent for the principal
eigenvalues of the covariance of the correlation
(Section \ref{sec:analytical}).

The mocks, and the covariance matrices of this paper, 
as well as covariance matrices with other binnings will
be available form the mocks website\footnote{http://www.marcmanera.net/mocks/}.

\section*{acknowledgments}
We thank the LasDamas collaboration for providing us with the simulation data used to calibrate our mocks.

MM acknowledges support from European Research Council.  

Funding for SDSS-III has been provided by the Alfred P. Sloan Foundation, the Participating Institutions, the National Science Foundation, and the U.S. Department of Energy Office of Science. The SDSS-III web site is http://www.sdss3.org/.

SDSS-III is managed by the Astrophysical Research Consortium for the
Participating Institutions of the SDSS-III Collaboration including the
University of Arizona,
the Brazilian Participation Group,
Brookhaven National Laboratory,
University of Cambridge,
Carnegie Mellon University,
University of Florida,
the French Participation Group,
the German Participation Group,
Harvard University,
the Instituto de Astrofisica de Canarias,
the Michigan State/Notre Dame/JINA Participation Group,
Johns Hopkins University,
Lawrence Berkeley National Laboratory,
Max Planck Institute for Astrophysics,
Max Planck Institute for Extraterrestrial Physics,
New Mexico State University,
New York University,
Ohio State University,
Pennsylvania State University,
University of Portsmouth,
Princeton University,
the Spanish Participation Group,
University of Tokyo,
University of Utah,
Vanderbilt University,
University of Virginia,
University of Washington,
and Yale University.

 The analysis made use of the computing resources of the
 National Energy Research Scientific Computing Center,
 the Shared Research Computing Services Pilot of the
 University of California and
 the Laboratory Research Computing project at
 Lawrence Berkeley National Laboratory.

 Part of the numerical computations and analyses were done on the Sciama High Performance Compute (HPC) cluster which is supported by the ICG, SEPNet and the University of Portsmouth. MM thanks Gary Burton and David Bacon for designing the system and for their help when running the mocks. 
 
 This research was partially supported by grant NSF AST-0607747

 We thank Molly Swanson for the courtesy of the plot of BOSS DR9 footprint.   


\begin{thebibliography}{99}


\bibitem[{{Abell et al.}(2009)}]{LSST}
Abell P.A., Allison J., Anderson S.F, Andrew J.R., Angel J.R.P., et al., 2009, arXiv 0912.0201

\bibitem[{{Anderson et al.}(2012)}]{boss12}
Anderson L. et al., 2012, submitted to MNRAS

\bibitem[{{Berlind \& Weinberg}(2002)}]{Wei02}
Berlind A.A., Weinberg D.H, 2002, ApJ, 575, 587-616

\bibitem[{{Bernardeau et al.}(2002)}]{Bernardeau02}
Bernardeau F., Colombi S., Gaztanaga E. and Scoccimarro R., 2002, PhR 367, Issue 1-3, p 1-248

\bibitem[{{Blake et al.}(2011)}]{Blake11}
Blake, C.; David, D.; Poole, G.G; Parkinson, D., 2011, MNRAS, 415, 2892-2909

\bibitem[{{Bouchet et al.}(1995)}]{bouchet}
Bouchet, F.R; Colombi, S.; Hivon, E.; Juszkiewicz, R.; AA 296 (1995) 575-608

\bibitem[{{Buchert}(1989)}]{Buc89}
Buchert T., 1989, A\&A, 223, 9

\bibitem[{{Bryan and Norman}(1998)}]{Bryan98}
Bryan G.L and Norman M.L, 1998, ApJ 4095, 80

\bibitem[{{Carlson \& White}(2010)}]{CarWhi10}
Carlson J., White M., 2010, ApJS, 190, 311

\bibitem[{{Cole et al.}(2005)}]{Cole05}
Cole S., Percival W.J, Peacock J.A., Norberg P., Baugh C., Frenk C.S., et al., 2005, MNRAS 362, Issue 2, p505-534

\bibitem[{{Coles et al.}(1993)}]{Coles93}
Coles, P., Melott, A.L., Shandarin, S.F., 1993, MNRAS 260, 765 

\bibitem[{{Croft et al.}(2011)}]{Croft11}
Croft R., Di Mateo T., Khandai N., Springel V., et al., 2011, arXiv1109.4169

\bibitem[{{Davis et al.}(1985)}]{DEFW}
 Davis M., Efstathiou G., Frenk C.S., White S.D.M., 1985, ApJ, 292, 371

\bibitem[{{Drinkwater et al.}(2010)}]{Drinkwater10}
Drinkwater M.J.; Jurek, R.J.; Blake C.; Woods D.; et al., 2010, MNRAS 401, 1429D

\bibitem[{{de Putter et al.}(2012)}]{deputtetal12}
de Putter, R., et al., 2012, arxiv:111.6596

\bibitem[{{Eisenstein et al.}(2011)}]{Eis11}
Eisenstein D.J., et al., 2011, AJ 142, 72

\bibitem[{{Feldman, Kaiser \& Peackok}(1994)}]{fkp94}
Feldman H.A., Kaiser N., Peacock J.A., 1994, ApJ, 426, 23

\bibitem[{{Fukugita et al.}(1996)}]{Fuk96}
Fukugita M., et al., 1996, AJ, 111, 1748

\bibitem[{{Giocoli et al.}(2010)}]{Gio10}
Giocoli C., Bartelmann M., Sheth R.K. and Caciato M., 2010, MNRAS 404, 502G

\bibitem[{{Gun et al.}(1998)}]{Gun98}
Gunn, J.E., et al. 1998, AJ, 116, 3040 (SDSS Camera)

\bibitem[{{Gun et al.}(2006)}]{Gun06}
Gunn, J.E., et al. 2006, AJ, 131, 2332 (SDSS Telescope)

\bibitem[{{Hamilton}(1992)}]{Ham92}
Hamilton A.J.S., 1992, ApJ, 385, L5-L8

\bibitem[{{Hamilton Rimes \& Scoccimarro}(2006)}]{Ham06}
Hamilton A.J.S., Rimes C.D., and Scoccimarro R., 2006, MNRAS 371, 1188

\bibitem[{{Hill et al.}(2004)}]{HETDEX}
Hill, G. J.; Gebhardt, K.;  Komatsu, E; MacQueen, P.J., The New Cosmology: Conference on Strings and Cosmology; The Mitchell Symposium on Observational Cosmology, AIP Conference Proc., 743, 2004, p.224-233

\bibitem[{{Hivon et al.}(1995)}]{Hiv95}
Hivon E., Bouchet F.R., Colombi S., Juszkiewicz R., 1995, A\&, 298, 643

\bibitem[{{Heitmann et al.}(2008)}]{Hei08}
Heitmann K., et al., 2008, CS\&D, 1, 15003

\bibitem[{{Kaiser} (1987)}]{kai87}
Kaiser, N., MNRAS 227 (1987) p1-21

\bibitem[{{Krewski}(1981)}]{Krewski81}
Krewski, D. and Rao, J.N.K.,1981, The Annals of Statistics. Vol. 9, 5, pp. 1010¿1019.

\bibitem[{{Kulkarni et al.}(2007)}]{Kulkarni07}
Kulkarni G.V., Nichol R.C., Sheth R.K., Seo H., Eisenstein D.J. and Gray A., 2007, MNRAS 378, 1196K


\bibitem[{{Larson et al.} (2011)}]{larsonwmap} 
Larson, D., Dunkley, J., Hinshaw, G., et al.\ 2011, ApJS, 192, 16

\bibitem[{{Landy \& Szalay}(1993)}]{lansza93}
Landy S.D., Szalay A.S., 1993, ApJ 412, 64

\bibitem[{{Lacey \& Cole}(1994)}]{Lacey94}
Lacey, C., \& Cole, S. 1994, MNRAS, 271, 676 

\bibitem[{{Laurejis et al.}(2011)}]{Euclid}
Laurejis R., Amiaux J., Ardunini S., Auguères J., et al., 2011, arXiv:1110.3193

\bibitem[{{Maccio et al.}(2008)}]{Maccio08}
Maccio A.V., Dutton A.A., and van den Bosch F.C., 2008, MNRAS, 391,1940

\bibitem[{{Matsubara}(2008a)}]{Mat08a}
 Matsubara T., 2008, Phys Rev D77, 063530

\bibitem[{{Matsubara}(2008b)}]{Mat08b}
 Matsubara T., 2008, Phys Rev D78, 083519

\bibitem[{{Moutarde et al.}(1991)}]{Mou91}
Moutarde F., Alimi J.-M., Bouchet F.R., Pellat R., Ramani A., 1991, ApJ, 382, 377

\bibitem[{{Navarro, Frenk \& White}(1996)}]{NFW}
Navarro, J., Frenk, C., White, S.D.M., 1996, ApJ, 462, 563

\bibitem[{{Nedler \& Mead}(1980)}]{nedler:1969}.
Nelder J.A. and Mead R., Computer Journal vol. 7 (1965), 308–313.

\bibitem[{{Norberg et al.}(2009)}]{Norberg09}
Norberg, P.; Baugh, C. M.; Gazta–aga, E. and Croton, D. J., MNRAS 396, 19, 2009). 

\bibitem[{{Nuza et al.}(2012)}]{nuza:2012}
Nuza S.E., Sanchez A.G., Prada F., Klypin D.J., et al. (2012) arXiv:12002.6057

\bibitem[{{Padmanabhan, White \& Cohn}(2009)}]{PadWhiCoh09}
Padmanabhan N., White M., Cohn J.D., 2009, Phys.Rev., D79, 063523

\bibitem[{{Padmanabhan \& White}(2009)}]{PadWhi09}
Padmanabhan N., White M., 2009, Phys. Rev., D80, 063508

\bibitem[{{Peacock \& Smith}(2000)}]{Peacock00}
Peacock J.A., Smith R.E., 200, MNRAS, 318,1144-1156

\bibitem[{{Percival et al.}(2010)}]{Percival10}
Percival W.J., Reid B.A., Eisenstein D.J., Bahcall N., et al., 2010, MNRAS, 401, 2148P

\bibitem[{{Pope and Smith}(2008)}]{Pope08}
Pope A.C. and Szapudi I., 2008, MNRAS 399, 766-774

\bibitem[{{Reid \& White}(2011)}]{Reid2011}
Reid B. and White M., 2011, MNRAS 417, 1913-1927

\bibitem[{{Reid et al.}(2012)}]{reid:2012}
Reid B., Samushia L., White M., et al., 2012, submitted to MNRAS 

\bibitem[{{Ross et al.}(2012)}]{Ross12}
Ross A.J., et al., 2012, submitted to MNRAS

\bibitem[{{Shao and Tu}(1995)}]{Shao95}
Shao, J. and Tu, D.,1995, The Jackknife and Bootstrap. Springer-Verlag, Inc.

\bibitem[{{Sheth et al.}(2001)}]{SMT01}
Sheth R.K., Mo H.J., \& Tormen G., 2001, MNRAS, 323, 1

\bibitem[{{Samushia et al.}(2012)}]{samshia:2012}
Samushia L., Reid B., White M., et. a., 2012, in prep 

\bibitem[{{Sanchez et al.}(2012)}]{sanchez:2012}
S\'{a}nchez A.G., Sc'{o}ccola C.G., Ross A. J., Percival W., et al., 2012, submitted to MNRAS 

\bibitem[{{Schlegel et al.}(2009a)}]{Boss2} 
Schlegel, D.~J., White M., and Eisenstein D., 2009, The Astronomy and Astrophysics Decadal Survey, Science White Papers \#314 arXiv:0902.4680

\bibitem[{{Schlegel et al.}(2009b)}]{BigBoss} 
Schlegel, D.~J., Bebek, C., Heetderks, H., et al.\ 2009, arXiv:0904.0468

\bibitem[{{Schneider et al.}(2011)}]{Schneider11} 
Schneider M.D., Frenk C.S. and Cole S., 2011, arXiv:111.5616

\bibitem[{{Scoccimarro} (1998)}]{Sco98}
Scoccimarro R., 1998, MNRAS 299, 1097

\bibitem[{{Scoccimarro et al.}(2001)}]{Sco01}
Scoccimarro R., Sheth R.K., Hui L., Jain B., 2001, ApJ 546, 20-34

\bibitem[{{Scoccimarro and Sheth}(2002)}]{Sco02}
Scoccimarro R., Sheth R.K, 2002, MNRAS 329, 629

\bibitem[{{Scoccimarro}(2004)}]{Scoccimarro04}
Scoccimarro, R., 2004, PRD 70, 083007

\bibitem[{{Scoccimarro}(2012)}]{Sco12}
Scoccimarro R., Hui L., Manera M., Chan L.C., 2012, arXiv: 11085.5512

\bibitem[{{Sefusatti \& Scoccimarro}(2005)}]{Sefusatti05}
Sefusatti E. and Scoccimarro R., 2006, PRD 71, 063001

\bibitem[{{Sefusatti et al.}(2006)}]{Sefusatti06}
Sefusatti E., Crocce M., Pueblas S. and Scoccimarro R., 2006, PRD 74, 023522 

\bibitem[{{Sefusatti et al.}(2012)}]{Sefusatti12}
Sefusatti E., Scoccimarro R., et al., 2012, in preparation

\bibitem[{{Sheth \& Tormen}(2004)}]{ShethT02}
Sheth R.K., Tormen G., 2004, MNRAS, 350, 1385

\bibitem[{{Skibba et al.}(2006)}]{Skibba06}
Skibba R., Sheth R.K., Connolly A.J. and Scranton R., 2006, MNRAS, 369, 68 

\bibitem[{{Skibba \& Sheth.}(2009)}]{Skibba09}
Skibba R. and Sheth R.K, 2009, MNRAS, 392, 1080

\bibitem[{{Smith \& Watts}(2005)}]{Smith05}
Smith R.E. and Watts P.I.R, 2005, MNRAS 360, 203


\bibitem[{{Smith, Sheth and Scoccimarro}(2008)}]{Smith08}
Smith R.E., Sheth R.E, Scoccimarro R., 2008, PRD 78, 023523 

\bibitem[{{Springel} (2005)}]{Spr05}
 Springel, V., MNRAS 2005, 364, 1105.

\bibitem[{{Swanson et al.}(2008)}]{Swanson08}
Swanson M.E.C., Tegmark M., Hamilton A.J.S. and  Hill J.C., 2008,  MNRAS 387, 1391

\bibitem[{{Tinker et al}(2008)}]{Tin08}
Tinker, J.L., Kravtsov, A.V., Klypin, A., Abazajian, K., Warren, M.S., Yepes, G., Gottloeber, S., Holz, D.E., 2008, ApJ, 688,  709


\bibitem[{{Taylor \& Hamilton}(1996)}]{TayHam96}
Taylor A.N., Hamilton A.J.S., 1996, MNRAS, 282, 767


\bibitem[{{Tojeiro et al.} (2012a)}]{tojeiro:2012a}
Tojeiro R., et al., 2012, submitted to MNRAS, arXiv:1202.6241

\bibitem[{{Tojeiro et al.} (2012b)}]{tojeiro:2012b}
Tojeiro R., et al., 2012, submitted to MNRAS

\bibitem[{{White et al.}(2010)}]{White10}
White M., Cohn J.D., Smith R. 2010, MNRAS, 408, 1818

\bibitem[{{White et al.} (2011)}]{white:2011}
White, Martin; Blanton, M.; Bolton, A.; Schlegel, D.; Tinker J.; et al., ApJ 728 (2011) Issue 2, 126

\bibitem[{{White}(2002)}]{TreePM}
White M., 2002, ApJS, 143, 241

\bibitem[{{Xu et al.}(2012)}]{Xu12}
Xu X., Padmanabhan N, Eisenstein D.J., Mehta K.T and Cuesta A.J, 2012, arXiv1202.0091

\bibitem[{{Zheng et al.}(2007)}]{Zhe07}
Zheng Z., Coil A., Zehavi I., 2007, ApJ, 667, 760








\end{thebibliography}

\clearpage
\begin{onecolumn}
\FloatBarrier

\begin{table}
\scalebox{0.8}{
\begin{tabular}{cccccccccccccccccccccccccccccccccccccccccccccccccccccccccccc}
\hline
\hline
$s (h^{-1})$Mpc& 30.8 & 33.4 & 36.2 & 39.2 & 42.5 & 46.1 & 49.9 & 54.1 & 58.7 & 63.6 & 68.9 & 74.7 & 81.0 & 87.8 & 95.1 \\
\hline
$\xi_0(s)$ NGC &497.07 & 408.22 & 333.79 & 270.19 & 216.45 & 171.90 & 134.90 & 104.60 & 79.56 & 59.34 & 43.06 & 30.87 & 22.05 & 17.49 &18.11 \\
\hline
$\xi_0(s)$ SGC & 484.07 & 397.71 & 322.96 & 261.15 & 207.05 & 163.71 & 127.21 & 96.39 & 72.85 & 53.02 & 37.46
& 25.43 & 17.12 & 12.90 & 13.25 \\
\hline
\hline
$s (h^{-1})$Mpc&  103.1 & 111.8 & 121.1 & 131.3 & 142.3 & 154.3 \\
\hline
$\xi_0(s)$ NGC & 19.39 & 15.45 & 6.84 & -0.45 & -3.77 & -4.43 \\
$\xi_0(s)$ SGC & 15.59 & 12.36 & 4.23 & -2.83 & -6.02 & -6.20 \\
\hline
\hline
\end{tabular}}
\caption{The monopole of the correlation function multiplied by $10^{5}$, for $\xi_0(s)$ with $30 < s < 160h^{-1}$Mpc for the DR9 NGC and SGC mocks. Correlation functions from PTHalos galaxy mock catalogues will be available with this and other binnings will be available at``http://www.marcmanera.net/mocks/''} 
\label{tablexi}
\end{table}

\begin{table}
\scalebox{0.8}{
\begin{tabular}{cccccccccccccccccccccccccccccccccccccccccccccccccccccccccccc}
\hline
\hline
$s (h^{-1})$Mpc& 30.8 & 33.4 & 36.2 & 39.2 & 42.5 & 46.1 & 49.9 & 54.1 & 58.7 & 63.6 & 68.9 & 74.7 & 81.0 & 87.8 & 95.1 &  103.1 & 111.8 & 121.1 & 131.3 & 142.3 & 154.3 \\
\hline
30.7 & 2.11 & 1.51 & 1.37 & 1.20 & 1.05 & 0.89 & 0.78 & 0.65 & 0.52 & 0.50 & 0.40 & 0.31 & 0.26 & 0.21 & 0.18 & 0.14 & 0.10 & 0.06 & 0.03 & 0.01 & -0.01 & \\ 
33.3 & 1.51 & 1.71 & 1.29 & 1.15 & 1.00 & 0.86 & 0.75 & 0.64 & 0.51 & 0.49 & 0.41 & 0.31 & 0.26 & 0.20 & 0.16 & 0.13 & 0.09 & 0.05 & 0.02 & -0.00 & -0.01 & \\ 
36.1 & 1.37 & 1.29 & 1.50 & 1.12 & 0.98 & 0.84 & 0.75 & 0.64 & 0.53 & 0.48 & 0.43 & 0.33 & 0.28 & 0.23 & 0.19 & 0.15 & 0.11 & 0.08 & 0.06 & 0.03 & 0.01 & \\ 
39.1 & 1.20 & 1.15 & 1.12 & 1.27 & 0.96 & 0.84 & 0.73 & 0.61 & 0.52 & 0.49 & 0.44 & 0.32 & 0.28 & 0.23 & 0.19 & 0.16 & 0.12 & 0.09 & 0.06 & 0.03 & 0.01 & \\ 
42.4 & 1.05 & 1.00 & 0.98 & 0.96 & 1.10 & 0.85 & 0.74 & 0.64 & 0.56 & 0.52 & 0.44 & 0.35 & 0.30 & 0.25 & 0.22 & 0.18 & 0.14 & 0.09 & 0.06 & 0.05 & 0.02 & \\ 
46.0 & 0.89 & 0.86 & 0.84 & 0.84 & 0.85 & 0.96 & 0.72 & 0.64 & 0.55 & 0.50 & 0.43 & 0.34 & 0.30 & 0.25 & 0.21 & 0.18 & 0.13 & 0.09 & 0.06 & 0.04 & 0.02 & \\ 
49.8 & 0.78 & 0.75 & 0.75 & 0.73 & 0.74 & 0.72 & 0.81 & 0.64 & 0.57 & 0.52 & 0.46 & 0.38 & 0.32 & 0.27 & 0.23 & 0.19 & 0.15 & 0.11 & 0.08 & 0.06 & 0.04 & \\ 
54.0 & 0.65 & 0.64 & 0.64 & 0.61 & 0.64 & 0.64 & 0.64 & 0.71 & 0.56 & 0.51 & 0.43 & 0.36 & 0.31 & 0.26 & 0.22 & 0.18 & 0.14 & 0.11 & 0.07 & 0.06 & 0.05 & \\ 
58.5 & 0.52 & 0.51 & 0.53 & 0.52 & 0.56 & 0.55 & 0.57 & 0.56 & 0.64 & 0.53 & 0.46 & 0.38 & 0.33 & 0.27 & 0.24 & 0.19 & 0.16 & 0.12 & 0.10 & 0.08 & 0.06 & \\ 
63.5 & 0.50 & 0.49 & 0.48 & 0.49 & 0.52 & 0.50 & 0.52 & 0.51 & 0.53 & 0.59 & 0.49 & 0.41 & 0.36 & 0.30 & 0.26 & 0.21 & 0.17 & 0.12 & 0.10 & 0.08 & 0.07 & \\ 
68.8 & 0.40 & 0.41 & 0.43 & 0.44 & 0.44 & 0.43 & 0.46 & 0.43 & 0.46 & 0.49 & 0.54 & 0.44 & 0.38 & 0.33 & 0.28 & 0.23 & 0.18 & 0.13 & 0.11 & 0.09 & 0.08 & \\ 
74.6 & 0.31 & 0.31 & 0.33 & 0.32 & 0.35 & 0.34 & 0.38 & 0.36 & 0.38 & 0.41 & 0.44 & 0.47 & 0.39 & 0.33 & 0.28 & 0.23 & 0.18 & 0.13 & 0.11 & 0.08 & 0.08 & \\ 
80.8 & 0.26 & 0.26 & 0.28 & 0.28 & 0.30 & 0.30 & 0.32 & 0.31 & 0.33 & 0.36 & 0.38 & 0.39 & 0.42 & 0.35 & 0.30 & 0.25 & 0.20 & 0.16 & 0.13 & 0.10 & 0.09 & \\ 
87.6 & 0.21 & 0.20 & 0.23 & 0.23 & 0.25 & 0.25 & 0.27 & 0.26 & 0.27 & 0.30 & 0.33 & 0.33 & 0.35 & 0.38 & 0.33 & 0.28 & 0.22 & 0.17 & 0.14 & 0.11 & 0.10 & \\ 
95.0 & 0.18 & 0.16 & 0.19 & 0.19 & 0.22 & 0.21 & 0.23 & 0.22 & 0.24 & 0.26 & 0.28 & 0.28 & 0.30 & 0.33 & 0.36 & 0.30 & 0.25 & 0.19 & 0.16 & 0.13 & 0.10 & \\ 
102.9 &  0.14 & 0.13 & 0.15 & 0.16 & 0.18 & 0.18 & 0.19 & 0.18 & 0.19 & 0.21 & 0.23 & 0.23 & 0.25 & 0.28 & 0.30 & 0.31 & 0.26 & 0.20 & 0.17 & 0.14 & 0.11 & \\ 
111.6 & 0.10 & 0.09 & 0.11 & 0.12 & 0.14 & 0.13 & 0.15 & 0.14 & 0.16 & 0.17 & 0.18 & 0.18 & 0.20 & 0.22 & 0.25 & 0.26 & 0.27 & 0.22 & 0.18 & 0.15 & 0.12 & \\ 
120.9 & 0.06 & 0.05 & 0.08 & 0.09 & 0.09 & 0.09 & 0.11 & 0.11 & 0.12 & 0.12 & 0.13 & 0.13 & 0.16 & 0.17 & 0.19 & 0.20 & 0.22 & 0.23 & 0.19 & 0.15 & 0.12 & \\ 
131.1 & 0.03 & 0.02 & 0.06 & 0.06 & 0.06 & 0.06 & 0.08 & 0.07 & 0.10 & 0.10 & 0.11 & 0.11 & 0.13 & 0.14 & 0.16 & 0.17 & 0.18 & 0.19 & 0.21 & 0.17 & 0.14 & \\ 
142.1 & 0.01 & -0.00 & 0.03 & 0.03 & 0.05 & 0.04 & 0.06 & 0.06 & 0.08 & 0.08 & 0.09 & 0.08 & 0.10 & 0.11 & 0.13 & 0.14 & 0.15 & 0.15 & 0.17 & 0.19 & 0.16 & \\ 
154.0 & -0.01 & -0.01 & 0.01 & 0.01 & 0.02 & 0.02 & 0.04 & 0.05 & 0.06 & 0.07 & 0.08 & 0.08 & 0.09 & 0.10 & 0.10 & 0.11 & 0.12 & 0.12 & 0.14 & 0.16 & 0.18 & \\ 
\hline
\label{tab:covn}
\end{tabular}}

\scalebox{0.8}{
\begin{tabular}{cccccccccccccccccccccccccccccccccccccccccccccccccccccccccccc}
\hline
\hline
$s (h^{-1})$Mpc& 30.7 & 33.3 & 36.1 & 39.1 & 42.4 & 46.0 & 49.8 & 54.0 & 58.5 & 63.5 & 68.8 & 74.6 & 80.8 & 87.6 & 95.0 & 102.9 & 111.6 & 120.9 & 131.1 & 142.1 & 154.0 & \\ 
\hline
30.7 & 6.80 & 5.11 & 4.36 & 3.95 & 3.31 & 2.94 & 2.57 & 2.08 & 1.75 & 1.36 & 1.19 & 0.83 & 0.74 & 0.64 & 0.63 & 0.46 & 0.33 & 0.27 & 0.24 & 0.15 & 0.08 & \\ 
33.3 & 5.11 & 6.11 & 4.44 & 3.87 & 3.40 & 3.12 & 2.79 & 2.25 & 1.87 & 1.48 & 1.22 & 0.91 & 0.79 & 0.69 & 0.60 & 0.46 & 0.38 & 0.30 & 0.28 & 0.22 & 0.14 & \\ 
36.1 & 4.36 & 4.44 & 5.08 & 3.88 & 3.42 & 3.03 & 2.65 & 2.18 & 1.83 & 1.43 & 1.17 & 0.91 & 0.78 & 0.70 & 0.59 & 0.46 & 0.42 & 0.34 & 0.30 & 0.24 & 0.19 & \\ 
39.1 & 3.95 & 3.87 & 3.88 & 4.46 & 3.51 & 3.07 & 2.66 & 2.23 & 1.89 & 1.49 & 1.16 & 0.92 & 0.77 & 0.68 & 0.57 & 0.43 & 0.34 & 0.31 & 0.26 & 0.20 & 0.17 & \\ 
42.4 & 3.31 & 3.40 & 3.42 & 3.51 & 4.06 & 3.18 & 2.77 & 2.30 & 1.95 & 1.48 & 1.20 & 1.00 & 0.85 & 0.70 & 0.58 & 0.46 & 0.38 & 0.32 & 0.29 & 0.20 & 0.17 & \\ 
46.0 & 2.94 & 3.12 & 3.03 & 3.07 & 3.18 & 3.70 & 2.92 & 2.46 & 2.04 & 1.65 & 1.42 & 1.17 & 0.93 & 0.81 & 0.63 & 0.46 & 0.39 & 0.33 & 0.30 & 0.21 & 0.18 & \\ 
49.8 & 2.57 & 2.79 & 2.65 & 2.66 & 2.77 & 2.92 & 3.24 & 2.45 & 2.09 & 1.70 & 1.44 & 1.18 & 0.99 & 0.83 & 0.67 & 0.50 & 0.41 & 0.34 & 0.31 & 0.24 & 0.21 & \\ 
54.0 & 2.08 & 2.25 & 2.18 & 2.23 & 2.30 & 2.46 & 2.45 & 2.76 & 2.14 & 1.77 & 1.48 & 1.21 & 1.00 & 0.83 & 0.65 & 0.52 & 0.43 & 0.39 & 0.37 & 0.29 & 0.24 & \\ 
58.5 & 1.75 & 1.87 & 1.83 & 1.89 & 1.95 & 2.04 & 2.09 & 2.14 & 2.36 & 1.83 & 1.53 & 1.23 & 1.06 & 0.87 & 0.69 & 0.55 & 0.44 & 0.40 & 0.36 & 0.29 & 0.26 & \\ 
63.5 & 1.36 & 1.48 & 1.43 & 1.49 & 1.48 & 1.65 & 1.70 & 1.77 & 1.83 & 1.99 & 1.58 & 1.28 & 1.11 & 0.92 & 0.77 & 0.64 & 0.50 & 0.47 & 0.42 & 0.33 & 0.28 & \\ 
68.8 & 1.19 & 1.22 & 1.17 & 1.16 & 1.20 & 1.42 & 1.44 & 1.48 & 1.53 & 1.58 & 1.80 & 1.39 & 1.17 & 0.98 & 0.80 & 0.65 & 0.52 & 0.46 & 0.42 & 0.33 & 0.27 & \\ 
74.6 & 0.83 & 0.91 & 0.91 & 0.92 & 1.00 & 1.17 & 1.18 & 1.21 & 1.23 & 1.28 & 1.39 & 1.52 & 1.21 & 1.00 & 0.81 & 0.66 & 0.51 & 0.42 & 0.39 & 0.31 & 0.25 & \\ 
80.8 & 0.74 & 0.79 & 0.78 & 0.77 & 0.85 & 0.93 & 0.99 & 1.00 & 1.06 & 1.11 & 1.17 & 1.21 & 1.36 & 1.12 & 0.91 & 0.76 & 0.60 & 0.49 & 0.43 & 0.36 & 0.30 & \\ 
87.6 & 0.64 & 0.69 & 0.70 & 0.68 & 0.70 & 0.81 & 0.83 & 0.83 & 0.87 & 0.92 & 0.98 & 1.00 & 1.12 & 1.29 & 1.04 & 0.88 & 0.71 & 0.58 & 0.48 & 0.40 & 0.34 & \\ 
95.0 & 0.63 & 0.60 & 0.59 & 0.57 & 0.58 & 0.63 & 0.67 & 0.65 & 0.69 & 0.77 & 0.80 & 0.81 & 0.91 & 1.04 & 1.16 & 0.98 & 0.82 & 0.66 & 0.55 & 0.45 & 0.38 & \\ 
102.9 & 0.46 & 0.46 & 0.46 & 0.43 & 0.46 & 0.46 & 0.50 & 0.52 & 0.55 & 0.64 & 0.65 & 0.66 & 0.76 & 0.88 & 0.98 & 1.11 & 0.92 & 0.75 & 0.63 & 0.51 & 0.44 & \\ 
111.6 & 0.33 & 0.38 & 0.42 & 0.34 & 0.38 & 0.39 & 0.41 & 0.43 & 0.44 & 0.50 & 0.52 & 0.51 & 0.60 & 0.71 & 0.82 & 0.92 & 1.03 & 0.83 & 0.69 & 0.56 & 0.47 & \\ 
120.9 & 0.27 & 0.30 & 0.34 & 0.31 & 0.32 & 0.33 & 0.34 & 0.39 & 0.40 & 0.47 & 0.46 & 0.42 & 0.49 & 0.58 & 0.66 & 0.75 & 0.83 & 0.88 & 0.74 & 0.59 & 0.48 & \\ 
131.1 & 0.24 & 0.28 & 0.30 & 0.26 & 0.29 & 0.30 & 0.31 & 0.37 & 0.36 & 0.42 & 0.42 & 0.39 & 0.43 & 0.48 & 0.55 & 0.63 & 0.69 & 0.74 & 0.84 & 0.69 & 0.56 & \\ 
142.1 & 0.15 & 0.22 & 0.24 & 0.20 & 0.20 & 0.21 & 0.24 & 0.29 & 0.29 & 0.33 & 0.33 & 0.31 & 0.36 & 0.40 & 0.45 & 0.51 & 0.56 & 0.59 & 0.69 & 0.74 & 0.62 & \\ 
154.0 & 0.08 & 0.14 & 0.19 & 0.17 & 0.17 & 0.18 & 0.21 & 0.24 & 0.26 & 0.28 & 0.27 & 0.25 & 0.30 & 0.34 & 0.38 & 0.44 & 0.47 & 0.48 & 0.56 & 0.62 & 0.72 & \\ 
\hline
\label{tab:covs}
\end{tabular}}
\caption{The covariance matrix, multiplied by $10^{5}$, for $\xi_0(s)$ with $30 < s < 160h^{-1}$Mpc, for the DR9 NGC (top) and SGC (bottom) footprint. Covariance matrices from PTHalo galaxy mock catalogues with this and other binnings will be available at ``http://www.marcmanera.net/mocks/''.} 
\end{table}

\end{onecolumn}
\end{document}